\documentclass[journal,twoside,web]{ieeecolor}
\usepackage{etoolbox}
\usepackage{color}
\makeatletter
\@ifundefined{color@begingroup}%
{\let\color@begingroup\relax
\let\color@endgroup\relax}{}%
\def\fix@ieeecolor@hbox#1{%
\hbox{\color@begingroup#1\color@endgroup}}
\patchcmd\@makecaption{\hbox}{\fix@ieeecolor@hbox}{}{\FAILED}
\patchcmd\@makecaption{\hbox}{\fix@ieeecolor@hbox}{}{\FAILED}
\usepackage{tmi}
\usepackage{bbding}
\usepackage{cite}
\usepackage{amsmath,amssymb,amsfonts}
\usepackage{bbm}
\usepackage{algorithmic}
\usepackage{graphicx}
\usepackage{textcomp}
\usepackage{algorithmic}
\usepackage{graphicx}
\usepackage{textcomp}
\usepackage{xcolor}
\usepackage{hyperref}
\definecolor{blueviolet}{rgb}{0.54, 0.17, 0.89}
\usepackage{graphicx}
\usepackage{booktabs} 
\usepackage{amssymb}
\usepackage{placeins}
\usepackage{xcolor}
\usepackage{amsmath}
\usepackage{colortbl}
\usepackage{colortbl}
\usepackage{amssymb}
\usepackage{multicol}
\usepackage{pifont}
\usepackage{multirow}
\usepackage{graphicx}
\usepackage[table,xcdraw]{xcolor}
\usepackage{makecell}

\colorlet{blue}{black}
\usepackage{bm}
\definecolor{right}{rgb}{0.99,0.001,0.001}
\definecolor{frenchblue}{rgb}{0.0, 0.45, 0.73}
\def\BibTeX{{\rm B\kern-.05em{\sc i\kern-.025em b}\kern-.08em
    T\kern-.1667em\lower.7ex\hbox{E}\kern-.125emX}}
\begin{document}
\title{Dictionary-based Pathology Mining with Hard-instance-assisted Classifier Debiasing for Genetic Biomarker Prediction from WSIs}
\author{Ling Zhang, Boxiang Yun, Ting Jin, Qingli Li, \IEEEmembership{Senior Member, IEEE}, Xinxing Li, and Yan Wang
\thanks{Manuscript submitted on Jan, 2025. This work was supported by the National Natural Science Foundation of China (Grant No. 62471182, 62101191), and Shanghai Rising-Star Program (Grant No. 24QA2702100), and the Science and Technology Commission of Shanghai Municipality (Grant No. 22DZ2229004).}
\thanks{L. Zhang, B. Yun, T. Jin, Q. Li, and Y. Wang are with Shanghai Key Laboratory of Multidimensional Information Processing, East China Normal University, Shanghai, China (e-mail: zhling@stu.ecnu.edu.cn, 52265904012@stu.ecnu.edu.cn, tjin@stu.ecnu.edu.cn, qlli@cs.ecnu.edu.cn, ywang@cee.ecnu.edu.cn).}
\thanks{X. Li is with Department of General Surgery, Tongji Hospital, Tongji University School of Medicine, Shanghai 200065, China (e-mail: ahtxxxx2015@163.com).}
\thanks{Corresponding author: Y. Wang}}

\maketitle

\begin{abstract}
Prediction of genetic biomarkers, \emph{e.g.}, microsatellite instability in colorectal cancer is crucial for clinical decision making. But, two primary challenges hamper accurate prediction: (1) It is difficult to construct a pathology-aware representation involving the complex interconnections among pathological components. (2) WSIs contain a large proportion of areas unrelated to genetic biomarkers, which make the model easily overfit simple but irrelative instances. We hereby propose a Dictionary-based hierarchical pathology mining with hard-instance-assisted classifier Debiasing framework to address these challenges, dubbed as D$^2$Bio. Our first module, dictionary-based hierarchical pathology mining, is able to mine diverse and very fine-grained pathological contextual interaction without the limit to the distances between patches. The second module, hard-instance-assisted classfier debiasing, learns a debiased classifier via focusing on hard but task-related features, without any additional annotations.  Experimental results on five cohorts show the superiority of our method, with over 4\% improvement in AUROC compared with the second best on the TCGA-CRC-MSI cohort. Our analysis further shows the clinical interpretability of D$^2$Bio in genetic biomarker diagnosis and potential clinical utility in survival analysis. \textcolor{blue}{Code will be available at \url{https://github.com/DeepMed-Lab-ECNU/D2Bio}}.

\end{abstract}


\section{Introduction}
\label{sec:introduction}
\IEEEPARstart{E}{valuation} of genetic biomarkers is crucial for cancer diagnosis and prognosis, such as microsatellite instability (MSI) and \textcolor{blue}{v-raf murine viral oncogene homolog B1 (BRAF)} for colorectal cancer and MSI for gastric cancer, since these biomarkers can identify patients with different treatment response and prognosis \cite{koncina2020prognostic,caputo2019braf,chang2018microsatellite,pietrantonio2019individual}. \textcolor{blue}{Specifically, MSI is an important biomarker for immunotherapy in colorectal and gastric cancers. In contrast, the BRAF mutation, present in about 10\% of colorectal cancer cases, is associated with poor prognosis and poor response to anti-EGFR therapies\cite{roth2010prognostic, di2008wild}.} Testing genetic biomarkers is time-consuming and expensive, with common methods, \emph{e.g.}, immunohistochemistry, \textcolor{blue}{polymerase chain reaction (PCR)}, and next-generation sequencing \cite{dedeurwaerdere2021comparison}. Due to diagnostic needs, whole slide images (WSIs) stained with hematoxylin and eosin (H\&E)  are routinely available for cancer patients. Besides, previous works \textcolor{blue}{\cite{kather2019deep, echle2020clinical, shimada2021histopathological}} suggest that genetic alterations are expressed in digital pathology WSIs. Therefore, automatically predicting genetic biomakers from WSIs is feasible and highly demanded in clinical practice.

\begin{figure}[t]
\centering
\includegraphics[width=0.45\textwidth]{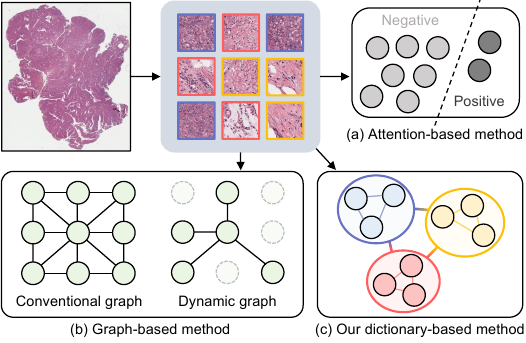}
\vspace{-0.6em}
    \caption{Motivation of our dictionary-based strategy.  (a) Attention-based MIL methods: simply model the relationships between instances. (b) Graph-based methods:  model instance relationships without construction of real pathological components. (c) Proposed dictionary-based method: utilizes a learnable dictionary to group instances into pathological components and hierarchically mine pathological contextual interaction.} 

\label{fig1}
\vspace{-1.6em}
\end{figure}
\textcolor{blue}{Genetic biomarker prediction from gigapixel WSIs is commonly formulated as a multiple instance learning (MIL) problem, in which only slide-level labels are available, without any patch-level annotations. In this setting, each WSI is treated as a bag, and the cropped patches extracted from the slide are regarded as instances, with the learning objective of predicting the corresponding slide-level label. Based on this paradigm, a wide range of MIL-based WSI classification methods have been developed, including attention-based methods \cite{Ilse_Tomczak_Welling_2018,li2021dual,lu2021data,shao2021transmil,li2021dt} and graph-based approaches \cite{chen2021whole,zheng2022graph,hou2023multi}.}
However, learning bag-level genetic biomarker representations and making accurate predictions is challenging. \textbf{First, constructing a pathology-aware representation including the complex interconnections among pathological components in the tumor micro-environment is difficult.} The tumor micro-environment includes a rich diversity of components, \emph{e.g.}, immune cells, cancer-associated fibroblasts (CAFs), endothelial cells (ECs), pericytes, and other cell types that vary by tissue—such
as adipocytes and neurons, with various pathological component interactions \cite{de2023evolving}. For genetic biomarker prediction, mining various pathological component interactions is of great importance \cite{bozyk2022tumor}. \textcolor{blue}{For example, in colorectal cancer, MSI arises from defects in the mismatch repair (MMR) system. This deficiency fails to correct errors occurring during DNA replication (particularly in repetitive microsatellite sequences), directly leading to the accumulation of somatic mutations and the subsequent generation of immunogenic neoantigens \cite{le2017mismatch}. This process triggers a robust host immune response, resulting in the extensive infiltration of Tumor-Infiltrating Lymphocytes (TILs) into the tumor microenvironment \cite{giannakis2016genomic}.} \textcolor{blue}{Similarly, BRAF mutation significantly remodels the tumor microenvironment through diverse pathological interactions. BRAF-mutated tumors, often arising via the serrated neoplasia pathway, are characterized by intense stromal remodeling and specific inflammatory responses \cite{debunne2013mucinous, kather2019deep}. Therefore, these complex spatial interactions among tumor cells, immune cells, and stromal components contain critical discriminative information.} Attention-based methods \cite{Ilse_Tomczak_Welling_2018,li2021dual,lu2021data,shao2021transmil,li2021dt} model the relationships between instances to distinguish positive from negative ones by simply merging multiple instance tokens (as depicted in Fig.~\ref{fig1} (a)). These methods struggle to model the complex tumor micro-environment without modeling diverse pathological components. Although HIPT \cite{chen2022scaling} designs a hierarchical transformer-based method, it is still restricted to exploring local pathological information in WSIs. Graph-based methods try to construct the tumor micro-environment (as depicted in Fig.~\ref{fig1} (b)). Conventional graph-based methods \cite{chen2021whole,zheng2022graph,hou2023multi} have demonstrated impressive results, but they construct the tumor micro-environment relying on fixed spatial positions, which limits the ability to explore mutual interaction in the tumor micro-environment freely. WiKG \cite{li2024dynamic} proposes a dynamic graph representation algorithm to enhance flexible interaction capabilities between instances at arbitrary locations. But it still struggles to represent real pathological components in the tumor micro-environment. 

\begin{figure}[t!]
\centering
\includegraphics[width=0.45\textwidth]{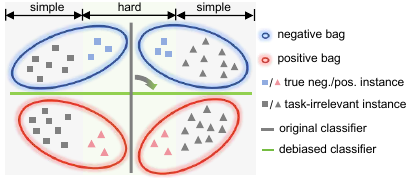}
\vspace{-0.3em}
    \caption{Motivation of our hard-instance-assisted classifier debiasing strategy. Sufferring from redundant task-irrelevant instances, during inference, biased classifier misjudges the bag-level labels by overfitting simple instances. Learning hard instances which include task-specific instances can assist the classifier to reduce the bias.}

\label{fig2}
\vspace{-0.6em}
\end{figure}

\textbf{Second, due to the nature of gigapixel WSIs, they typically contain a large proportion of areas unrelated to genetic biomarkers. While the interconnections between pathological components are learned, the pathology-aware representation still inevitably acquires information that is irrelevant to genetic biomarkers.} 
Extracting genetic biomarkers-related features is like finding a needle in a haystack. The large proportion of irrelevant regions in a WSI 
makes the model easily overfit simple but irrelative instances, ignoring hard but task-specific instances (as depicted in Fig.~\ref{fig2}). To solve this problem, instance-level interventions are introduced. \cite{lu2021data} assigns attention-based pseudo labels to several instances to further supervise the model, making the model focus on simple instances and easily overfit task-irrelevant features. \cite{tang2023multiple}  highlights the attention-based ``hard-to-classify" instances by masking simple instances, leading to the lost of important information. How to enable the model to focus on hard but biomarker classification-relevant features without losing important information is non-trivial. 

To address these issues, we propose a \textbf{D}ictionary-based hierarchical pathology mining with hard-instance-assisted classifier \textbf{D}ebiasing method for WSI-based genetic \textbf{Bio}marker prediction, dubbed as D$^2$Bio. D$^2$Bio consists of two modules: (1) dictionary-based hierarchical pathology mining, which learns a pathology-aware representation of WSIs and (2) hard-instance-assisted classifier debiasing, which learns a debiased classifier via focusing on hard but task-related features. Concretely, inspired by sparse dictionary representation \cite{kreutz2003dictionary}, which aims to represent a signal using fewer elements, we introduce a learnable dictionary to construct fine-grained pathological components based on real pathological information in complex WSIs. After constructing pathological components, the hierarchical contextual interaction of intra- and inter-pathological components are extracted and fused (as depicted in Fig.~\ref{fig1} (c)). Compared with attention-based methods \cite{Ilse_Tomczak_Welling_2018,shao2021transmil} and graph-based methods \cite{chen2021whole,li2024dynamic}, our dictionary-based hierarchical pathology mining enjoys the benefits of completely breaking the limit of instance locations to model the real pathology distribution and extracting hierarchical interaction among fine-grained pathological components in WSIs. Furthermore, we propose a hard-instance-assisted classifier debiasing module to prevent the classifier from overfitting irrelative simple information in WSIs. Specifically, we focus on hard instances in WSIs, and propose an unsupervised clustering strategy to separate positive and negative instances as pseudo labels for training the classifier. Through the dictionary-based hierarchical pathology mining module, a pathology-aware representation of the WSI is constructed, providing meaningful pathology information to the classifier. Then the classifier enhances its ability to focus on relevant features through the hard-instance-assisted classifier debiasing module. Thus, we extract comprehensive and in-depth biomarker-associated features in WSIs and make robust decisions for genetic biomarker prediction. Extensive experiments show that our method outperforms all state-of-the-arts. Additionally, our method shows clinical interpretability, providing potential for future clinical applications.
\begin{figure*}[t!]
\centering
    \includegraphics[width=0.95\textwidth]{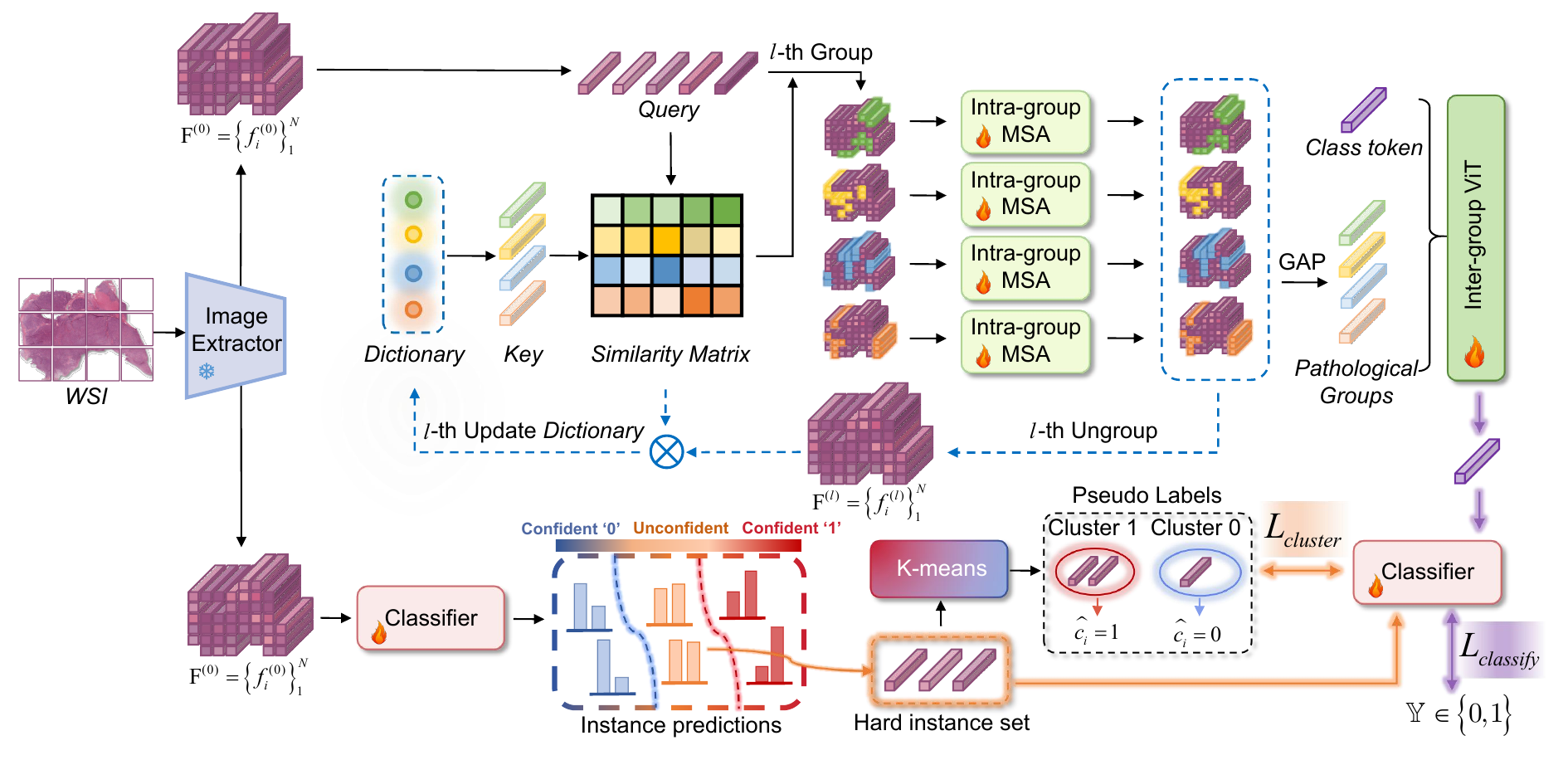}
    \vspace{-0.6em}
    \caption{Illustration of D$^2$Bio. The overall framework consists of two parts: 1) dictionary-based hierarchical pathology mining, 2) hard-instance assisted-classifier debiasing. Given a WSI $\mathbf{X}$, our D$^2$Bio first initialize a learnable dictionary to extract pathological information of $\mathbf{X}$ via cross attention operation. Then our D$^2$Bio groups instances into fine-grained pathological groups according to the similarity matrix. To hierarchically mine the interaction, Multi-head Self-attention (MSA) is first employed in each group. Features of $\mathbf{X}$ are updated by ungrouping these groups, which is further used to update the dictionary. After repeating the above steps $L$ times, inter-group ViT is employed. Finally, our D$^2$Bio assigns hard instance pseudo labels via unsupervised clustering to supervise the classfication head to reduce the bias in WSIs. The \textcolor{blueviolet}{purple arrow} indicates the classification branch and the \textcolor{orange}{orange arrow} indicates the classifier debiasing branch.}
\label{fig3}
\vspace{-0.6em}
\end{figure*}
The contributions of our work are summarized as follows: 
\begin{itemize}
\item  We propose a dictionary-based hierarchical pathology mining module which extracts very fine-grained pathological components and explores complex interconnections in WSIs without distance limitation between patches.
\item We design a hard-instance-assisted classifier debiasing module to learn a robust classifier, preventing the classifier from overfitting irrelative redundant information in WSIs, without any additional annotations.
\item Extensive experiments on five cohorts demonstrate that our method significantly outperforms mainstream MIL models (\emph{e.g.}, with over 4\% improvement in AUROC on the TCGA-CRC-MSI cohort). Our proposed modules can be easily plugged into other MIL methods, offering interpretability and potential usage for survival analysis. 
\end{itemize}

The preliminary version of D$^2$Bio, PromptBio \cite{zhang2024prompting}, was presented as a conference paper at the 27th International Conference on Medical Image Computing and Computer Assisted Intervention (MICCAI 2024). This extended work brings significant improvements: (1) we utilize a learnable dictionary to guide instance grouping instead of prompting large language models; (2) we learn a debiased classifier via focusing on hard but task-related features and (3) we add two new cohorts to evaluate the effectiveness of D$^2$Bio, supplemented with extra experiments, ablation studies and clinical analysis. 

\section{Related Work}
\subsection{MIL Methods for WSI Classification}
Multiple instance learning (MIL) is widely used in whole-slide image (WSI) classification, where WSIs are treated as bags and image patches as instances. MIL methods can be broadly categorized into instance-based and bag-based approaches. Instance-based methods \cite{qu2022dgmil,campanella2019clinical,chikontwe2020multiple} generate instance predictions and aggregate them to obtain the bag prediction, while bag-based methods \cite{wagner2023transformer,hashimoto2020multi,yao2020whole,li2021dual,zhang2022dtfd} focus on extracting instance features to build bag representations for training a classifier. In bag-based approaches, attention mechanisms \cite{Ilse_Tomczak_Welling_2018,li2021dual,lu2021data} are commonly used to identify key instances within WSIs. These methods aim to find a \textcolor{blue}{``}needle in a haystack\textcolor{blue}{''} as the positive region in WSIs is typically small, and they treat instances as independent entities, overlooking their contextual interactions. More recently, transformer-based methods \cite{shao2021transmil,li2021dt} have been proposed to model inter-instance relationships, but these methods often merge instance tokens without considering hierarchical contextual interactions within WSIs. Several studies \cite{lu2021data,shi2020loss,qu2022bi,qu2024rethinking,wang2024rethinking} have demonstrated the benefits of instance-level supervision for improving bag-level classification performance. For example, \cite{lu2021data} assigns pseudo-labels to all instances in negative bags and to instances with the highest and lowest attention scores in positive bags. \cite{wang2024rethinking} utilizes bag-level classifiers to generate pseudo-labels for instances, fine-tuning the patch embedder to enhance bag-level classification. In \cite{tang2023multiple}, hard instances are emphasized by masking simple ones to mitigate bias towards them. In this paper, we focus on hierarchically modeling interactions between diverse pathological components in WSIs and reducing bias of the classifier by unsupervised pseudo-labeling of hard instances.
\subsection{Sparse Dictionary Learning}
Signal processing techniques commonly require sparse representations, which capture the useful characteristics of the signal. Sparse representation of a signal involves the choice of a dictionary, which is the set of atoms used to decompose the signal \cite{kreutz2003dictionary}. Recently sparse dictionary learning has been successfully applied to image analysis, such as image recognition \cite{lu2017simultaneous,tang2020dictionary}, denoising \cite{li2012group,zheng2021deep} and super-resolution \cite{jiang2018super,ayas2020single,zhang2024transcending}. \textcolor{blue}{In the domain of computational pathology, dictionary learning has also been utilized for stain normalization \cite{vahadane2016structure} and tissue classification using discriminative dictionaries built on handcrafted features \cite{vu2015histopathological, srinivas2014simultaneous}.} In WSIs, there \textcolor{blue}{are} diverse cellular tissue structures, ranging from coarse-grained (lymphocytes, inflammatory cells, adenocarcinoma, \emph{etc.}) to fine-grained (B cells, neutrophils, cancer cells, \emph{etc.}). Within the fine-grained pathological components, the morphology and size of cells are similar. These both lead to sparsity of  gigapixel WSIs. Inspired by sparse dictionary representation, we consider gigapixel WSIs as sparse signals, which can be effectively represented by a dictionary composed of fine-grained pathological components, leading to a dense representation of WSIs. \textcolor{blue}{However, unlike prior pathology methods that construct discriminative dictionaries based on handcrafted features  and utilizes them primarily for feature reconstruction or sparse coding, our method integrates a learnable dictionary within an end-to-end deep learning framework to extract fine-grained pathological components in WSIs, leading to a pathology-aware representation of WSIs.}

\section{Methodology}

Mathematically, given a WSI $\mathbf{X}\in\mathbb{R}^{W\times H\times 3}$, whose size is $W\times H$, the goal of our task is to predict the image-level label $\hat{\mathbbm{Y}}=\{0,1\}$. 
Our D$^2$Bio consists of two modules: 1) dictionary-based hierarchical pathology mining and 2) hard-instance-assisted classifier debiasing. In the first module, two steps are included, \emph{i.e.}, dictionary-guided instance grouping and hierarchical interaction mining. Specifically, we first initialize a learnable dictionary $\mathbf{D}$ to extract the pathological information of $\mathbf{X}$ via cross-attention operation. A large number of atoms in the dictionary can ensure that the dictionary extracts sufficiently fine-grained pathological information from the WSI. Then we group instances into fine-grained pathological groups according to the similarity between each instance in $\mathbf{X}$ and each atom in $\mathbf{D}$. To mine the interaction within each pathological group, Multi-head Self-Attention (MSA) in each group is first employed. The feature of $\mathbf{X}$ is then updated after ungrouping these pathological groups and is subsequently used to update the dictionary.  After repeating the above steps $L$ times, we obtain pathological groups, which are termed as pathological components. Then inter-group Vision Transformer (ViT) is employed to mine the interaction among the pathological groups. This dictionary-based hierarchical pathology mining strategy benefits the modeling of a complex and diverse environment in WSIs. To learn a debiased binary classifier for differentiating between positive and negative instances, we further assign hard instance pseudo labels via unsupervised clustering, which is utilized to assist the classification head to identify important underlying entities that are difficult to discern.
\subsection{Problem Formulation}
In a binary classification task based on the multiple instance learning setting, a WSI is considered as a bag and the patches cropped from the WSI are regarded as instances. The bag is negative only when all instances are negative, and the bag is positive when at least one instance is positive. We consider the prediction of genetic biomarkers based on WSIs as a bag-level classification task, where only the bag label $\mathbbm{Y}\in\{0,1\}$ is available. The feature of the $i$-th instance is obtained through feeding the $i$-th cropped patch into a frozen image extractor, denoted as $\mathbf{f}_i^{(0)}\in\mathbb{R}^d$, where $d$ is the dimension of the feature.
\subsection{Dictionary-based Hierarchical Pathology Mining}
\subsubsection{Dictionary-Guided Instance Grouping}
Sparse dictionary representation \cite{kreutz2003dictionary} is a form of representation learning which originated in signal processing. It aims to represent a signal using fewer elements, and through dictionary learning, it can effectively enhance the sparsity of the signal. The assumption of dictionary learning is that given a signal set $\mathbf{S}=\{\mathbf{s}_{i}\}_{i=1}^{\dot{N}}\in \mathbb{R}^{\dot{d} \times \dot{N}}$, where $\dot{d}$ is the dimension of a single signal and $\dot{N}$ is the number of signals, we hope to find a dictionary $\dot{\mathbf{D}}=\left[\dot{\mathbf{d}}_1, \dot{\mathbf{d}}_2, \ldots, \dot{\mathbf{d}}_{\dot{K}}\right] \in \mathbb{R}^{\dot{d} \times \dot{K}}$, where each of the dictionary's columns is an atom and $\dot{K}$ is the number of atoms, and an expression $\mathbf{E}\in \mathbb{R}^{\dot{K}\times \dot{N}}$, subject to $\mathbf{S} \simeq \dot{\mathbf{D}} \mathbf{E}$, to seek the sparse representation.  When $\dot{K}<\dot{d}$, the dictionary is an undercomplete dictionary. More precisely, a sparse dictionary problem can be viewed as the following optimization problem:
\begin{equation}
\underset{\dot{\mathbf{D}} \in \mathbf{\mathcal{D}}, \mathbf{E} \in \mathbf{\mathcal{E}}}{\min } \left\|\mathbf{S} - \dot{\mathbf{D}}\mathbf{E}\right\|_F^2,
\end{equation}
where $\left\| \cdot \right\|_F$ means the Frobenius norm, and $\mathbf{\mathcal{D}}$ and $\mathbf{\mathcal{E}}$ are possible sets of the dictionaries and expressions.

Most sparse dictionary learning methods perform a two-stage procedure iteratively to choose the dictionary: start with an initial dictionary and repeat the following two stages (\emph{e.g.}, Eq.~\eqref{eq2} and Eq.~\eqref{eq3}) several times. Stage 1 uses the current dictionary to find the sparse expression and stage 2 updates the dictionary to reduce the error of stage 1.

\noindent \textbf{Stage 1} Sparse representation:
\begin{equation}
\label{eq2}
    \mathbf{E}^{(l+1)}=\underset{\mathbf{E} \in \mathcal{E}}{\mathrm{argmin}}\left\|\mathbf{S}-\dot{\mathbf{D}} ^{(l)} \mathbf{E}^{(l)}\right\|_{F}^{2}.
\end{equation}
\textbf{Stage 2} Dictionary updating:
\begin{equation}
\label{eq3}
\dot{\mathbf{D}} ^{(l+1)}=\underset{\dot{\mathbf{D}} \in \mathcal{D}}{\mathrm{argmin}}\left\|\mathbf{S}-\dot{\mathbf{D}}^{(l)}  \mathbf{E}^{(l+1)}\right\|_{F}^{2}.
\end{equation}
where $l$ is the number of updates.
Inspired by sparse dictionary learning, we consider a WSI as an input sparse signal, which contains substantial instances. We establish a dictionary to densely represent the WSI and extract pathological components in this section. We first initialize a dictionary $\mathbf{D}^{(0)}=\left[\mathbf{d}_1^{(0)}, \mathbf{d}_2^{(0)}, \ldots, \mathbf{d}_K^{(0)}\right] \in \mathbb{R}^{K \times d}$, which includes $K$ atoms of dimension $d$. To densely represent the WSI, we set $K<d$. We aim to learn a specific and comprehensive representation of the WSI using the dictionary. Therefore, we treat the WSI features $\mathbf{F}^{(0)}\in\mathbb{R}^{N\times d}$, consisting of $N$ instances and the $i$-th instance feature is $\mathbf{f}_i^{(0)}$, as the query, and treat the initialized dictionary $\mathbf{D}^{(0)}$ as the key to extract pathological information in the WSI:
\begin{equation}
\widetilde{\mathbf{Q}}^{(0)}= \mathbf{F}^{(0)} \cdot \mathbf{W}_{Q},~~~ \widetilde{\mathbf{K}}^{(0)}=\mathbf{D}^{(0)} \cdot \mathbf{W}_{K},
\end{equation}
\noindent{\textcolor{blue}{where $\mathbf{W}_q \in \mathbb{R}^{d \times r}$ and $\mathbf{W}_k \in \mathbb{R}^{d \times r}$ project inputs to a lower dimension ($r=4$) for efficient attention computation.}}

Then we calculate the similarity between $\widetilde{\mathbf{Q}}^{(0)}$ and $\widetilde{\mathbf{K}}^{(0)}$:
\begin{equation}
\mathbf{H}^{(0)}=\mathrm{Softmax}\left(\mathrm{sim}\left(\widetilde{\mathbf{Q}}^{(0)}, \widetilde{\mathbf{K}}^{(0)}\right) / \gamma\right),
\end{equation}
where $\mathrm{sim}(\textit{·}\textit{ , }\textit{·})$ refers to cosine similarity and $\gamma$ is a parameter to adjust the similarity range.

This gives the similarity $\mathbf{H}^{(0)}\in \mathbb{R}^{N \times K}$ between each instance feature in $\mathbf{F}^{(0)}$ and each atom in the dictionary $\mathbf{D}^{(0)}$.


Then we aim to construct explicit pathological groups to model the diverse fine-grained pathological components in WSIs. We group instances into  $K$ pathological components, \emph{i.e.}, $\{\mathbf{P}^{(0)}_{k}\}_{k=1}^{K}$, based on the similarity $\mathbf{H}^{(0)}$ (see Eq. \eqref{eq6}). Specifically, if $h^{(0)}_{i k}$ is the maximum similarity in $\{h^{(0)}_{i 1}, h^{(0)}_{i 2}, \ldots, h^{(0)}_{i K}\}$, where $h^{(0)}_{i k}$ is the similarity between the $i$-th instance feature $\mathbf{f}^{(0)}_{i}$ and the $k$-th atom in $\mathbf{D}^{(0)}$, then the instance feature $\mathbf{f}^{(0)}_{i}$ is assigned to the $k$-th group $\mathbf{P}^{(0)}_{k}$:
\begin{equation}
\label{eq6}
\mathbf{P}^{(0)}_{k}=\left\{\mathbf{f}^{(0)}_{i} \mid \arg\underset{j}\max \left(h^{(0)}_{i j}\right)=k\right\}_{i=1}^N,
\end{equation}
where $k,j\in \{1,2,...,K\}$.

\begin{table*}[t]
\centering
\normalsize
\setlength{\tabcolsep}{1pt} 
\renewcommand{\arraystretch}{1.2} 

\caption{\textcolor{blue}{Comparisons with state-of-the-art MIL methods on five cohorts using two feature extractors, \emph{i.e.}, CtransPath \cite{wang2022transformer} and UNI v2 \cite{chen2024towards}. We report Mean (95\% CI) in percentage (\%). ``*'' denotes statistical significance (p $<$ 0.05). Best results are shown in \textbf{bold}.}}
\label{tab:main_results_compact}

\resizebox{\textwidth}{!}{
\begin{tabular}{l|cccc|cccc|cccc|cccc|cccc}
\toprule[1.5pt] 
\multicolumn{21}{c}{\textbf{CtransPath \cite{wang2022transformer}}} \\ \hline
\multirow{2}{*}{Method} & \multicolumn{4}{c|}{TCGA-CRC-MSI} & \multicolumn{4}{c|}{TCGA-CRC-BRAF} & \multicolumn{4}{c|}{CPTAC-MSI} & \multicolumn{4}{c|}{CPTAC-BRAF} & \multicolumn{4}{c}{TCGA-STAD-MSI} \\
\cline{2-21}
 & AUC & Acc & PRC & Bal & AUC & Acc & PRC & Bal & AUC & Acc & PRC & Bal & AUC & Acc & PRC & Bal & AUC & Acc & PRC & Bal \\
\hline 

%
\multirow{2}{*}{ABMIL \cite{Ilse_Tomczak_Welling_2018}} 
& 88.00$^{*}$ & 80.80$^{*}$ & 64.80 & \textbf{86.20} 
& 68.40$^{*}$ & 80.80 & 24.80 & 54.90$^{*}$ 
& 87.50 & 76.20$^{*}$ & 49.10 & 50.00$^{*}$ 
& 85.20 & 85.70$^{*}$ & 56.90 & 50.00$^{*}$ 
& 77.80 & 65.70$^{*}$ & 46.80$^{*}$ & 67.80$^{*}$ \\[-1ex]
& \scriptsize{(75.5-96.6)} & \scriptsize{(73.3-87.5)} & \scriptsize{(46.0-83.1)} & \scriptsize{(77.6-92.4)} 
& \scriptsize{(46.1-87.6)} & \scriptsize{(74.2-87.5)} & \scriptsize{(6.4-55.7)} & \scriptsize{(41.5-71.0)} 
& \scriptsize{(68.8-100.0)} & \scriptsize{(76.2-76.2)} & \scriptsize{(29.8-100.0)} & \scriptsize{(50.0-50.0)} 
& \scriptsize{(72.6-99.4)} & \scriptsize{(85.7-85.7)} & \scriptsize{(21.8-95.8)} & \scriptsize{(50.0-50.0)} 
& \scriptsize{(67.8-87.2)} & \scriptsize{(56.6-74.7)} & \scriptsize{(35.2-68.2)} & \scriptsize{(57.7-77.8)}
\\

\multirow{2}{*}{DSMIL \cite{li2021dual}} 
& 88.20$^{*}$ & 87.50$^{*}$ & 62.70 & 81.40 
& 63.80$^{*}$ & 76.70$^{*}$ & 24.70 & 52.70$^{*}$ 
& 91.20 & 76.20 & 57.30 & 56.90$^{*}$ 
& 87.00 & 85.70 & 38.80 & 50.00 
& 77.10 & 63.60$^{*}$ & 47.70$^{*}$ & 63.80$^{*}$ \\[-1ex]
& \scriptsize{(76.1-97.5)} & \scriptsize{(81.7-93.3)} & \scriptsize{(40.2-87.9)} & \scriptsize{(68.6-92.4)} 
& \scriptsize{(40.5-83.8)} & \scriptsize{(70.0-83.3)} & \scriptsize{(5.6-56.8)} & \scriptsize{(39.3-69.2)} 
& \scriptsize{(76.2-100.0)} & \scriptsize{(61.9-85.7)} & \scriptsize{(34.4-100.0)} & \scriptsize{(40.6-76.9)} 
& \scriptsize{(78.1-92.2)} & \scriptsize{(85.7-85.7)} & \scriptsize{(22.1-75.2)} & \scriptsize{(50.0-50.0)} 
& \scriptsize{(66.7-86.6)} & \scriptsize{(54.5-72.7)} & \scriptsize{(35.2-69.3)} & \scriptsize{(53.1-73.8)}
\\

\multirow{2}{*}{CLAM-SB \cite{lu2021data}} 
& 85.80$^{*}$ & 88.30 & 69.80 & 84.80 
& 70.00$^{*}$ & 76.70$^{*}$ & 33.60 & 64.30 
& 90.00 & 81.00 & 55.50 & 80.60 
& 77.80 & 85.70$^{*}$ & 50.10 & 63.90$^{*}$ 
& 77.60 & 68.70$^{*}$ & 47.70$^{*}$ & 69.80$^{*}$ \\[-1ex]
& \scriptsize{(72.0-97.1)} & \scriptsize{(82.5-94.2)} & \scriptsize{(47.3-88.1)} & \scriptsize{(73.8-94.3)} 
& \scriptsize{(48.9-88.2)} & \scriptsize{(69.2-84.2)} & \scriptsize{(6.8-62.3)} & \scriptsize{(46.9-83.0)} 
& \scriptsize{(73.8-100.0)} & \scriptsize{(61.9-95.2)} & \scriptsize{(33.0-100.0)} & \scriptsize{(60.0-96.9)} 
& \scriptsize{(70.4-88.9)} & \scriptsize{(81.4-90.5)} & \scriptsize{(31.6-73.7)} & \scriptsize{(48.9-79.2)} 
& \scriptsize{(67.4-87.2)} & \scriptsize{(59.6-77.8)} & \scriptsize{(35.8-68.6)} & \scriptsize{(59.1-79.8)}
\\

\multirow{2}{*}{TransMIL \cite{shao2021transmil}} 
& 83.70$^{*}$ & 80.00$^{*}$ & 52.20$^{*}$ & 77.10 
& 62.60$^{*}$ & 70.00$^{*}$ & 26.60 & 60.70 
& 88.70 & 81.00 & 55.80 & 80.60 
& 92.60 & 85.70$^{*}$ & 67.80 & 50.00$^{*}$ 
& 67.90$^{*}$ & 61.60$^{*}$ & 47.30 & 62.40$^{*}$ \\[-1ex]
& \scriptsize{(70.4-94.3)} & \scriptsize{(73.3-86.7)} & \scriptsize{(30.2-79.0)} & \scriptsize{(65.2-88.1)} 
& \scriptsize{(38.9-85.3)} & \scriptsize{(61.7-78.3)} & \scriptsize{(5.7-58.8)} & \scriptsize{(42.9-78.6)} 
& \scriptsize{(71.3-100.0)} & \scriptsize{(61.9-95.2)} & \scriptsize{(32.1-100.0)} & \scriptsize{(57.5-96.9)} 
& \scriptsize{(78.9-99.3)} & \scriptsize{(85.7-85.7)} & \scriptsize{(36.4-96.4)} & \scriptsize{(50.0-50.0)} 
& \scriptsize{(56.5-79.2)} & \scriptsize{(52.5-71.7)} & \scriptsize{(31.8-63.8)} & \scriptsize{(51.7-73.1)}
 \\

\multirow{2}{*}{DTFD \cite{zhang2022dtfd}} 
& 84.19$^{*}$ & 85.00$^{*}$ & 58.47$^{*}$ & 77.15 
& 70.54$^{*}$ & 79.17$^{*}$ & 26.79 & 54.02$^{*}$ 
& 87.50$^{*}$ & 76.20$^{*}$ & 59.26$^{*}$ & 50.00$^{*}$ 
& 87.00$^{*}$ & 85.70$^{*}$ & 62.50$^{*}$ & 50.00$^{*}$ 
& 75.40$^{*}$ & 65.70$^{*}$ & 45.60$^{*}$ & 70.40 \\[-1ex]
& \scriptsize{(72.6-95.8)} & \scriptsize{(78.3-90.8)} & \scriptsize{(34.2-79.6)} & \scriptsize{(63.8-89.5)} 
& \scriptsize{(52.4-86.9)} & \scriptsize{(72.5-85.8)} & \scriptsize{(7.1-55.4)} & \scriptsize{(40.2-70.1)} 
& \scriptsize{(70.0-100.0)} & \scriptsize{(76.2-76.2)} & \scriptsize{(29.6-100.0)} & \scriptsize{(50.0-50.0)} 
& \scriptsize{(70.4-88.7)} & \scriptsize{(85.7-85.7)} & \scriptsize{(46.1-73.7)} & \scriptsize{(50.0-50.0)} 
& \scriptsize{(64.7-84.8)} & \scriptsize{(56.6-74.7)} & \scriptsize{(34.1-66.5)} & \scriptsize{(60.4-79.1)}
 \\

\multirow{2}{*}{HIPT \cite{chen2022scaling}} 
& 83.90$^{*}$ & 74.20$^{*}$ & 61.80 & 76.70 
& 72.80 & 68.30$^{*}$ & \textbf{40.00} & 71.40 
& 88.80 & 76.20$^{*}$ & 67.10 & 70.60$^{*}$ 
& 92.60 & 81.00$^{*}$ & 67.80 & 75.00 
& 79.60 & 62.60$^{*}$ & 54.20 & 72.40 \\[-1ex]
& \scriptsize{(70.4-94.7)} & \scriptsize{(66.7-81.7)} & \scriptsize{(38.7-81.6)} & \scriptsize{(65.7-86.2)} 
& \scriptsize{(46.0-95.5)} & \scriptsize{(60.0-76.7)} & \scriptsize{(10.4-74.8)} & \scriptsize{(54.9-85.7)} 
& \scriptsize{(71.3-100.0)} & \scriptsize{(57.1-90.5)} & \scriptsize{(32.3-100.0)} & \scriptsize{(46.8-93.8)} 
& \scriptsize{(87.4-99.6)} & \scriptsize{(76.7-90.0)} & \scriptsize{(36.2-97.6)} & \scriptsize{(60.0-91.7)} 
& \scriptsize{(69.7-88.7)} & \scriptsize{(53.5-70.7)} & \scriptsize{(39.6-75.7)} & \scriptsize{(64.3-79.7)}

 \\

\multirow{2}{*}{MHIM-MIL \cite{tang2023multiple}} 
& 83.00$^{*}$ & 85.83$^{*}$ & 56.42$^{*}$ & 77.62 & 74.90 & 76.67$^{*}$ & 24.50 & 75.89 & 92.50 & 81.00 & 72.45 & 80.63 & 88.05 & 85.71$^{*}$ & 59.24 & 77.78 & 66.11$^{*}$ & 49.49$^{*}$ & 34.56$^{*}$ & 56.95$^{*}$ \\[-1ex]
& \scriptsize{(71.5-94.6)} & \scriptsize{(79.2-91.7)} & \scriptsize{(32.2-81.1)} & \scriptsize{(64.3-89.3)} & \scriptsize{(56.1-93.7)} & \scriptsize{(68.3-84.2)} & \scriptsize{(08.5-52.2)} & \scriptsize{(54.2-91.0)} & \scriptsize{(80.8-100)} & \scriptsize{(61.9-95.2)} & \scriptsize{(35.4-100)} & \scriptsize{(60.6-96.9)} & \scriptsize{(66.2-92.6)} & \scriptsize{(66.7-95.2)} & \scriptsize{(18.5-100)} & \scriptsize{(52.8-94.4)} & \scriptsize{(54.2-78.0)} & \scriptsize{(39.4-59.6)} & \scriptsize{(21.1-51.2)} & \scriptsize{(46.8-67.1)} \\

\multirow{2}{*}{WiKG \cite{li2024dynamic}} 
& 90.20$^{*}$ & 90.00 & 67.30 & 82.90 
& 80.20 & 67.50$^{*}$ & 17.20 & \textbf{76.80} 
& 92.50 & 90.50 & 77.50 & \textbf{93.80} 
& 92.60 & 85.70 & 65.60 & \textbf{91.70} 
& 79.80 & 70.70$^{*}$ & 54.10 & 64.50$^{*}$ \\[-1ex]
& \scriptsize{(81.5-97.2)} & \scriptsize{(84.2-95.0)} & \scriptsize{(43.8-89.5)} & \scriptsize{(70.5-93.3)} 
& \scriptsize{(67.9-90.7)} & \scriptsize{(59.2-75.8)} & \scriptsize{(9.6-51.3)} & \scriptsize{(62.1-86.6)} 
& \scriptsize{(77.5-100.0)} & \scriptsize{(76.2-100.0)} & \scriptsize{(39.7-100.0)} & \scriptsize{(84.4-100.0)} 
& \scriptsize{(93.1-100.0)} & \scriptsize{(81.4-95.2)} & \scriptsize{(80.5-100.0)} & \scriptsize{(89.2-97.2)} 
& \scriptsize{(70.3-88.4)} & \scriptsize{(62.6-79.8)} & \scriptsize{(39.2-76.5)} & \scriptsize{(53.8-75.2)}

 \\

\multirow{2}{*}{PromptBio \cite{zhang2024prompting}} 
& 91.80$^{*}$ & 86.70$^{*}$ & 45.90$^{*}$ & 63.80$^{*}$ 
& 80.10 & 70.00$^{*}$ & 30.90 & 72.30 
& 93.80 & 90.50 & 62.70 & 86.90 
& 92.60 & 71.40$^{*}$ & 65.60 & 83.30 
& 77.00 & 73.70 & 43.60$^{*}$ & 66.50$^{*}$ \\[-1ex]
& \scriptsize{(86.1-96.1)} & \scriptsize{(81.7-91.7)} & \scriptsize{(31.9-73.4)} & \scriptsize{(51.4-76.7)} 
& \scriptsize{(61.3-94.1)} & \scriptsize{(61.7-78.3)} & \scriptsize{(10.2-62.1)} & \scriptsize{(54.9-86.2)} 
& \scriptsize{(81.2-100.0)} & \scriptsize{(76.2-100.0)} & \scriptsize{(38.7-100.0)} & \scriptsize{(66.9-100.0)} 
& \scriptsize{(79.3-96.1)} & \scriptsize{(48.6-80.0)} & \scriptsize{(25.2-83.1)} & \scriptsize{(70.0-88.3)} 
& \scriptsize{(67.2-86.3)} & \scriptsize{(65.7-81.8)} & \scriptsize{(32.9-62.6)} & \scriptsize{(55.9-77.2)}

 \\

\multirow{2}{*}{\textbf{D$^2$Bio (ours)}} 
& \textbf{96.70} & \textbf{92.50} & \textbf{72.20} & 75.70 
& \textbf{83.80} & \textbf{85.80} & 20.90 & 63.40 
& \textbf{96.30} & \textbf{90.50} & \textbf{86.00} & 86.90 
& \textbf{94.40} & \textbf{95.20} & \textbf{81.70} & 83.30 
& \textbf{81.70} & \textbf{77.80} & \textbf{59.00} & \textbf{77.20} \\[-1ex]
& \scriptsize{(93.1-99.4)} & \scriptsize{(88.3-95.8)} & \scriptsize{(50.7-97.1)} & \scriptsize{(62.9-88.6)} 
& \scriptsize{(74.3-92.4)} & \scriptsize{(80.0-90.8)} & \scriptsize{(11.4-54.6)} & \scriptsize{(46.0-81.7)} 
& \scriptsize{(85.0-100.0)} & \scriptsize{(76.2-100.0)} & \scriptsize{(48.3-100.0)} & \scriptsize{(66.9-100.0)} 
& \scriptsize{(76.1-100.0)} & \scriptsize{(91.0-100.0)} & \scriptsize{(50.9-100.0)} & \scriptsize{(68.3-100.0)} 
& \scriptsize{(71.9-90.6)} & \scriptsize{(69.7-85.9)} & \scriptsize{(42.7-77.7)} & \scriptsize{(67.2-86.5)}
\\
\hline
\multicolumn{21}{c}{\textbf{UNI v2 \cite{chen2024towards}}} \\ \hline
\multirow{2}{*}{Method} & \multicolumn{4}{c|}{TCGA-CRC-MSI} & \multicolumn{4}{c|}{TCGA-CRC-BRAF} & \multicolumn{4}{c|}{CPTAC-MSI} & \multicolumn{4}{c|}{CPTAC-BRAF} & \multicolumn{4}{c}{TCGA-STAD-MSI} \\
\cline{2-21}
 & AUC & Acc & PRC & Bal & AUC & Acc & PRC & Bal & AUC & Acc & PRC & Bal & AUC & Acc & PRC & Bal & AUC & Acc & PRC & Bal \\
\hline 

\multirow{2}{*}{ABMIL \cite{Ilse_Tomczak_Welling_2018}} 
& 93.30$^{*}$ & 90.00$^{*}$ & 63.90$^{*}$ & 82.90 
& 88.30$^{*}$ & 92.50 & 38.60$^{*}$ & 72.80 
& 93.80$^{*}$ & \textbf{90.50}$^{*}$ & 62.70$^{*}$ & 86.90$^{*}$ 
& 94.10 & 85.70 & 73.10 & 81.60 
& 81.60 & 68.70$^{*}$ & 56.10 & 71.10$^{*}$ \\[-1ex]
& \scriptsize{(86.8-98.2)} & \scriptsize{(84.2-95.0)} & \scriptsize{(43.3-91.8)} & \scriptsize{(70.5-93.8)} 
& \scriptsize{(69.6-93.9)} & \scriptsize{(92.5-94.8)} & \scriptsize{(20.8-46.2)} & \scriptsize{(55.9-78.1)} 
& \scriptsize{(76.2-93.1)} & \scriptsize{(72.9-90.0)} & \scriptsize{(35.6-58.3)} & \scriptsize{(69.1-86.6)} 
& \scriptsize{(94.3-100.0)} & \scriptsize{(81.9-99.5)} & \scriptsize{(76.5-100.0)} & \scriptsize{(70.7-99.7)} 
& \scriptsize{(71.8-90.1)} & \scriptsize{(59.6-77.8)} & \scriptsize{(40.7-75.9)} & \scriptsize{(61.0-79.8)}
\\
%

\multirow{2}{*}{DSMIL \cite{li2021dual}} 
& 89.20$^{*}$ & 90.80$^{*}$ & 58.90$^{*}$ & 71.90$^{*}$
& 73.10$^{*}$ & 75.80$^{*}$ & 38.30 & 63.80$^{*}$
& 93.80 & \textbf{90.50} & 62.70 & 86.90
& 95.60 & \textbf{90.50} & 83.50 & 84.60$^{*}$
& 84.10 & 74.70$^{*}$ & 61.40 & 76.50 \\[-1ex]
& \scriptsize{(78.7-96.5)} & \scriptsize{(86.7-95.0)} & \scriptsize{(37.6-82.5)} & \scriptsize{(59.0-85.2)}
& \scriptsize{(46.6-84.1)} & \scriptsize{(67.1-79.7)} & \scriptsize{(14.2-53.7)} & \scriptsize{(47.6-71.4)}
& \scriptsize{(88.1-99.4)} & \scriptsize{(81.9-94.8)} & \scriptsize{(50.0-96.2)} & \scriptsize{(74.4-95.9)}
& \scriptsize{(92.8-99.4)} & \scriptsize{(85.7-94.3)} & \scriptsize{(77.4-98.7)} & \scriptsize{(81.6-86.9)}
& \scriptsize{(73.4-93.1)} & \scriptsize{(66.6-82.8)} & \scriptsize{(46.2-85.2)} & \scriptsize{(66.4-85.2)}
 \\

\multirow{2}{*}{CLAM-SB \cite{lu2021data}} 
& 94.00$^{*}$ & 83.30$^{*}$ & 87.60 & 71.50$^{*}$
& 83.80 & 85.00$^{*}$ & 74.60 & 28.40
& 95.00 & \textbf{90.50} &\textbf{ 93.80} & 77.70
& 91.20$^{*}$ & 85.70 & 81.60 & 51.90$^{*}$
& 78.60$^{*}$ & 68.70$^{*}$ & 71.10$^{*}$ & 57.10$^{*}$ \\[-1ex]
& \scriptsize{(86.7-98.9)} & \scriptsize{(76.7-90.0)} & \scriptsize{(79.5-93.8)} & \scriptsize{(49.4-93.4)}
& \scriptsize{(78.3-98.2)} & \scriptsize{(78.6-89.9)} & \scriptsize{(61.2-93.3)} & \scriptsize{(19.2-67.1)}
& \scriptsize{(88.2-99.6)} & \scriptsize{(86.2-99.5)} & \scriptsize{(90.9-99.7)} & \scriptsize{(51.7-98.6)}
& \scriptsize{(88.2-98.4)} & \scriptsize{(81.0-94.8)} & \scriptsize{(61.5-93.5)} & \scriptsize{(42.6-94.0)}
& \scriptsize{(68.1-88.1)} & \scriptsize{(59.6-77.8)} & \scriptsize{(61.1-81.1)} & \scriptsize{(40.7-73.5)}
 \\

\multirow{2}{*}{TransMIL \cite{shao2021transmil}} 
& 90.20$^{*}$ & 82.50$^{*}$ & 84.30$^{*}$ & 48.20$^{*}$
& 77.60$^{*}$ & 81.70$^{*}$ & 72.80 & 36.30
& 95.00 & 81.00 & 66.90$^{*}$ & 77.70
& 89.70 & 76.20$^{*}$ & 75.70$^{*}$ & 50.10
& 84.50 & 80.80 & 73.90 & 55.60$^{*}$ \\[-1ex]
& \scriptsize{(82.5-96.3)} & \scriptsize{(75.0-89.2)} & \scriptsize{(74.3-92.4)} & \scriptsize{(32.4-78.3)}
& \scriptsize{(61.5-87.9)} & \scriptsize{(76.2-88.1)} & \scriptsize{(49.4-86.3)} & \scriptsize{(19.9-69.7)}
& \scriptsize{(95.5-100.0)} & \scriptsize{(76.7-90.0)} & \scriptsize{(57.2-79.0)} & \scriptsize{(79.6-100.0)}
& \scriptsize{(88.8-100.0)} & \scriptsize{(76.7-90.0)} & \scriptsize{(79.3-93.8)} & \scriptsize{(42.8-100.0)}
& \scriptsize{(76.1-91.8)} & \scriptsize{(72.7-87.9)} & \scriptsize{(63.8-84.6)} & \scriptsize{(41.6-76.8)}
 \\

\multirow{2}{*}{DTFD \cite{zhang2022dtfd}} 
& 95.90$^{*}$ & 93.30 & \textbf{90.50} & 72.60$^{*}$
& 84.30$^{*}$ & 85.80$^{*}$ & \textbf{80.80} & 42.90
& 96.30 & \textbf{90.50} & 86.90 & 86.00
& 85.30$^{*}$ & 81.00$^{*}$ & 88.20$^{*}$ & 39.90$^{*}$
& 82.80 & 75.80$^{*}$ & 75.80 & 65.50 \\[-1ex]
& \scriptsize{(90.3-99.6)} & \scriptsize{(89.2-97.5)} & \scriptsize{(80.5-98.1)} & \scriptsize{(52.9-98.0)}
& \scriptsize{(60.4-87.0)} & \scriptsize{(81.1-86.6)} & \scriptsize{(62.8-86.5)} & \scriptsize{(21.3-59.1)}
& \scriptsize{(89.1-99.7)} & \scriptsize{(77.1-99.5)} & \scriptsize{(71.9-99.7)} & \scriptsize{(72.1-99.2)}
& \scriptsize{(76.6-89.7)} & \scriptsize{(67.1-85.2)} & \scriptsize{(79.7-90.9)} & \scriptsize{(29.3-45.9)}
& \scriptsize{(72.9-91.2)} & \scriptsize{(66.7-83.8)} & \scriptsize{(65.2-85.8)} & \scriptsize{(49.2-80.7)}
 \\

\multirow{2}{*}{HIPT \cite{chen2022scaling}} 
& 94.60$^{*}$ & 87.50$^{*}$ & 87.10$^{*}$ & 67.40$^{*}$
& 77.30$^{*}$ & 82.50$^{*}$ & 73.20 & 27.80
& 93.80$^{*}$ & \textbf{90.50} & \textbf{93.80} & 62.70$^{*}$
& 91.20 & 85.70 & 91.20 & 51.90
& 78.70$^{*}$ & 77.80$^{*}$ & \textbf{77.20} & 51.60$^{*}$ \\[-1ex]
& \scriptsize{(89.3-98.5)} & \scriptsize{(81.7-93.3)} & \scriptsize{(77.1-95.2)} & \scriptsize{(45.4-91.0)}
& \scriptsize{(56.5-92.9)} & \scriptsize{(80.2-85.0)} & \scriptsize{(66.2-85.6)} & \scriptsize{(18.1-61.9)}
& \scriptsize{(75.0-93.1)} & \scriptsize{(76.7-90.0)} & \scriptsize{(84.7-93.4)} & \scriptsize{(33.9-58.1)}
& \scriptsize{(79.3-100.0)} & \scriptsize{(72.9-90.5)} & \scriptsize{(83.2-94.1)} & \scriptsize{(33.0-100.0)}
& \scriptsize{(68.4-87.7)} & \scriptsize{(69.7-85.9)} & \scriptsize{(67.1-86.5)} & \scriptsize{(37.0-67.9)}
\\

\multirow{2}{*}{MHIM-MIL \cite{tang2023multiple}} 
& 94.90$^{*}$ & 86.70$^{*}$ & 49.50$^{*}$ & 63.90$^{*}$
& 82.00$^{*}$ & 92.50 & 61.20 & 25.50$^{*}$
& 95.00 & 85.70 & 76.90$^{*}$ & 77.70
& 82.40 & 81.00 & 50.00$^{*}$ & 80.60
& 79.50 & 75.80$^{*}$ & 67.90$^{*}$ & 51.00$^{*}$ \\[-1ex]
& \scriptsize{(88.8-98.9)} & \scriptsize{(85.0-87.5)} & \scriptsize{(48.6-50.0)} & \scriptsize{(44.9-92.8)}
& \scriptsize{(79.4-92.4)} & \scriptsize{(90.1-91.7)} & \scriptsize{(54.1-65.9)} & \scriptsize{(17.9-30.7)}
& \scriptsize{(92.6-100.0)} & \scriptsize{(85.7-94.8)} & \scriptsize{(71.4-89.7)} & \scriptsize{(51.2-100.0)}
& \scriptsize{(65.0-100.0)} & \scriptsize{(81.0-81.0)} & \scriptsize{(50.0-50.0)} & \scriptsize{(59.9-100.0)}
& \scriptsize{(70.0-88.6)} & \scriptsize{(67.7-83.8)} & \scriptsize{(57.2-78.6)} & \scriptsize{(37.7-72.2)}
\\

\multirow{2}{*}{WiKG \cite{li2024dynamic}} 
& 95.50$^{*}$ & 94.20 & 82.40$^{*}$ & 67.10$^{*}$
& 85.00$^{*}$ & 90.80$^{*}$ & 54.50$^{*}$ & 25.10$^{*}$
& 96.30 & 71.40$^{*}$ & 81.20$^{*}$ & 86.00
& 94.10 & 90.50 & \textbf{94.10} & 57.90
& 75.40$^{*}$ & 75.80 & 73.20 & 47.10$^{*}$ \\[-1ex]
& \scriptsize{(90.6-98.9)} & \scriptsize{(90.0-97.5)} & \scriptsize{(69.5-93.3)} & \scriptsize{(46.0-94.2)}
& \scriptsize{(79.4-94.5)} & \scriptsize{(86.2-93.2)} & \scriptsize{(46.8-60.2)} & \scriptsize{(17.8-40.1)}
& \scriptsize{(90.0-100.0)} & \scriptsize{(62.9-76.2)} & \scriptsize{(75.6-84.4)} & \scriptsize{(63.9-100.0)}
& \scriptsize{(83.5-100.0)} & \scriptsize{(86.2-100.0)} & \scriptsize{(91.5-100.0)} & \scriptsize{(37.7-100.0)}
& \scriptsize{(64.9-85.5)} & \scriptsize{(67.7-83.8)} & \scriptsize{(63.1-83.2)} & \scriptsize{(34.0-65.3)}
 \\

\multirow{2}{*}{PromptBio \cite{zhang2024prompting}} 
& 96.80 & 91.70$^{*}$ & 75.20$^{*}$ & 69.20
& 90.20$^{*}$ & 72.50$^{*}$ & 79.50 & 31.80
& 95.00 & 76.20$^{*}$ & 50.00$^{*}$ & 77.70
& 85.30$^{*}$ & 81.00 & 78.70$^{*}$ & 42.30$^{*}$
& 81.40$^{*}$ & 77.80 & 67.90$^{*}$ & 50.50$^{*}$ \\[-1ex]
& \scriptsize{(93.4-99.2)} & \scriptsize{(87.5-95.8)} & \scriptsize{(62.4-88.6)} & \scriptsize{(49.0-95.4)}
& \scriptsize{(81.6-93.1)} & \scriptsize{(69.3-74.1)} & \scriptsize{(67.1-85.7)} & \scriptsize{(21.7-52.4)}
& \scriptsize{(93.9-100.0)} & \scriptsize{(76.2-76.2)} & \scriptsize{(50.0-50.0)} & \scriptsize{(64.0-100.0)}
& \scriptsize{(69.3-95.1)} & \scriptsize{(67.1-90.5)} & \scriptsize{(63.9-84.6)} & \scriptsize{(24.6-83.7)}
& \scriptsize{(71.5-90.0)} & \scriptsize{(69.7-84.8)} & \scriptsize{(57.2-78.0)} & \scriptsize{(37.3-72.1)}
 \\

\multirow{2}{*}{\textbf{D$^2$Bio (ours)}} 
& \textbf{98.50} & \textbf{96.70} & 87.50 & \textbf{92.40}
& \textbf{92.60} & \textbf{94.20} & 52.60 & \textbf{73.70}
& \textbf{97.50} & \textbf{90.50} & \textbf{93.80} & \textbf{93.80}
& \textbf{97.10} & \textbf{90.50} & 87.10 & \textbf{94.10}
& \textbf{86.20} & \textbf{82.80} & 70.90 & \textbf{80.60} \\[-1ex]
& \scriptsize{(96.2-100.0)} & \scriptsize{(93.3-99.2)} & \scriptsize{(69.0-100.0)} & \scriptsize{(82.4-99.5)}
& \scriptsize{(87.7-94.2)} & \scriptsize{(94.2-95.8)} & \scriptsize{(43.6-61.4)} & \scriptsize{(68.8-79.0)}
& \scriptsize{(96.4-100.0)} & \scriptsize{(85.7-94.8)} & \scriptsize{(92.1-100.0)} & \scriptsize{(90.6-96.6)}
& \scriptsize{(94.1-100.0)} & \scriptsize{(77.1-90.5)} & \scriptsize{(44.0-100.0)} & \scriptsize{(85.9-94.1)}
& \scriptsize{(76.1-93.7)} & \scriptsize{(74.7-89.9)} & \scriptsize{(54.1-85.5)} & \scriptsize{(70.5-88.6)}
\\
\bottomrule[1pt]
\end{tabular}
}
\end{table*}

In this way, each instance can be assigned to the corresponding pathological group based on its similarity to the atoms in the dictionary. But, this grouping process results in varying numbers of instances in each pathological group, making parallel computation difficult. To obtain balanced pathological groups, all pathological groups $\{\mathbf{P}^{(0)}_{k}\}_{k=1}^{K}$ are flattened, concatenated and redivided into $G$ pathological groups, \emph{i.e.}, $\{\dot{\mathbf{P}}^{(0)}_{g}\}_{g=1}^{G}$, where each pathological group has the same number of instances. In the above steps, we record the position information for each instance. In this way, we group all instances of the WSI into $G$ pathological components based on the learnable dictionary.





After the grouping, we propose an intra-group interaction mining strategy, whose details will be introduced in Sec.~\ref{method:hierarchical}. Then we ungroup all pathological components to update the WSI features, \emph{i.e.}, $\mathbf{F}^{(1)}$, based on the recorded position information for each instance. Following sparse dictionary learning, we use  $\mathbf{F}^{(1)}$ and the similarity $\mathbf{H}^{(0)}$ to update the dictionary, followed by setting a parameter $\tau$ for further momentum updating the dictionary $\mathbf{D}^{(0)}$. \textcolor{blue}{It is important to note that before this ungrouping, the grouped features have already been processed by the Intra-group MSA (detailed in Sec. III-B-2) and aggregated via Global Average Pooling (GAP) to generate features of pathological components. These features are preserved for the subsequent Inter-group ViT, ensuring that the semantic structure is fully leveraged rather than discarded, while ungrouping allows the instance features to be realigned with the similarity matrix for dictionary refinement.} Thus, we obtain an updated dictionary $\mathbf{D}^{(1)}$. We repeat the above process for $L$ times, which can be formulated as below:
\begin{equation}
\label{eq11}
\begin{array}{c}
\hat{\mathbf{D}}^{(l)}=\mathrm{Softmax} \left(\mathrm{Norm}\left({\mathbf{H}^{(l)}}^\top\right)\right)\cdot \mathbf{F}^{(l)} ,
\end{array}
\end{equation}
\begin{equation}
\label{eq12}
\begin{array}{c}
\mathbf{D}^{(l+1)}=\tau \hat{\mathbf{D}}^{(l)}+(1-\tau) \mathbf{D}^{(l)},
\end{array}
\end{equation}
\noindent where $l\in \{1,2,...,L\}$, $\mathrm{Softmax}(\cdot)$ and $\mathrm{Norm}(\cdot)$ indicate softmax and instance norm operations, respectively. $\tau$ is a learnable parameter initialized to 0.
\begin{figure}[t!]
\centering
\includegraphics[width=0.46\textwidth]{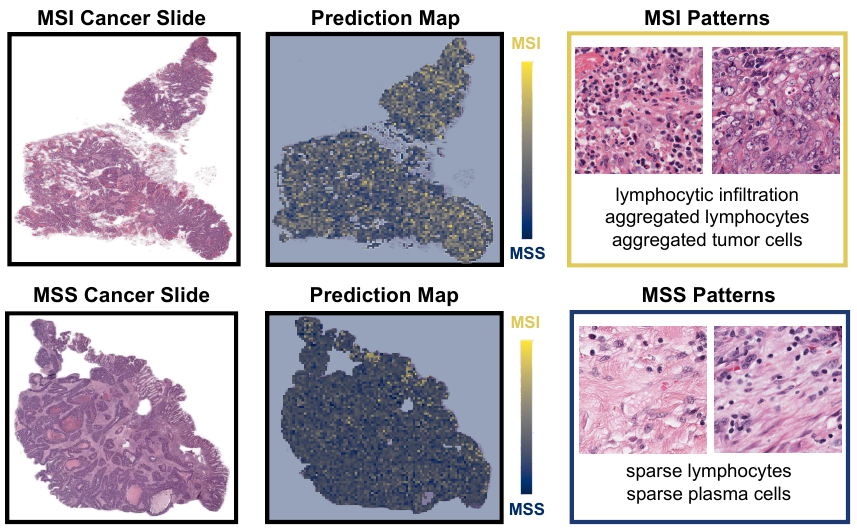}
\vspace{-0.6em}
\caption{Heatmap visualization of D$^2$Bio on the MSI prediction task and identified pathological patterns.
} \label{fig5}
\vspace{-1.6em}
\end{figure}
\subsubsection{Hierarchical Interaction Mining}
\label{method:hierarchical}
In this section, we introduce hierarchically mining contextual interaction of intra- and inter-pathological components. For interaction of intra-pathological components, various interaction occur among lymphocytes to regulate immune responses, forming a complex immune network. 
As for interaction of inter-pathological components, lymphocytes and inflammatory cells interact through signaling molecules and chemical mediators to regulate the intensity of immune responses. Therefore, mining interaction of intra- and inter-pathological components counts a lot for modeling the tumor micro-environment. 


To mine the interaction within each fine-grained pathological component $\dot{\mathbf{P}}^{(l)}_g$, where $g=\{1,2,...,G\}$, we conduct intra-group MSA operation on all the instance features in $\dot{\mathbf{P}}^{(l)}_g$:
\begin{equation}
\ddot{\mathbf{P}}^{(l)}_g=\mathrm{MSA}\left(\dot{\mathbf{P}}^{(l)}_g \cdot \mathbf{W}_q, \dot{\mathbf{P}}^{(l)}_g \cdot \mathbf{W}_k, \dot{\mathbf{P}}^{(l)}_g \cdot \mathbf{W}_v\right),
\end{equation}
\noindent \textcolor{blue}{where $\mathbf{W}_q \in \mathbb{R}^{d \times r}$ and $\mathbf{W}_k \in \mathbb{R}^{d \times r}$ project inputs to a lower dimension ($r=4$) to compute attention scores efficiently, while $\mathbf{W}_v \in \mathbb{R}^{d \times d}$ maintains the full feature dimension.}

After conducting MSA operation within each fine-grained pathological component and updating the dictionary $L$ times, we separately aggregate the instance features of each pathological group by global average pooling and obtain all the features of pathological component, \emph{i.e.}, $\{\bar{\mathbf{P}}_g\}_{g=1}^{G}$.

Then, we concatenate a class token  ${\mathbf{CLS}}\in \mathbb{R}^{d}$ with all the group features. The concatenated feature is further fed into the inter-group ViT layer to mine the interaction among all the pathological groups. We use a Multi-Layer Perception (MLP) head to map the output class token  ${\mathbf{CLS}}^{\prime} \in \mathbb{R}^{d}$ to the final class probability $\hat{\mathbbm{P}}$. So far, we design a simple yet effective dictionary-based hierarchical pathology mining strategy.

We use the genetic biomarker label $\mathbbm{Y}\in\{0,1\}$ to supervise the classification head. The cross-entropy loss is adopted for calculating the classification loss:
\begin{equation}
\label{eq17}
\mathcal{L}_{cls}=\mathrm{CELoss}\left(\hat{\mathbbm{P}},\mathbbm{Y}\right).
\end{equation}

\subsection{Hard-Instance-Assisted Classifier Debiasing}
Due to the complexity and diversity of biological phenomena, many phenotypic features in WSIs are not absolutely specific, leading to inherent bias in the dataset. With the bias, the model easily overfits simple instances and relies on non-specific features, leading to misclassification. Therefore, instead of only learning simple instances, we focus on learning hard instances, identifying underlying patterns that are difficult to discern but truly relevant to the label. Specifically,  we feed all the original instance features $\mathbf{F}^{(0)}=\{\mathbf{f}^{(0)}_1,\mathbf{f}^{(0)}_2,\ldots,\mathbf{f}^{(0)}_N \}$ into the final classification head $\mathcal{C}$, outputting the prediction logits of each instance $\hat{p}_i$ 
as follows:
\begin{equation}
\hat{p}_i=\mathrm{Sigmoid}\left(\mathcal{C}\left(\mathbf{f}^{(0)}_i\right)\right), i \in \{1,2,\ldots,N\},
\label{eq10}
\end{equation}
where $\mathrm{Sigmoid}(\cdot)$ means the sigmoid operation.

If the difference between $\hat{p}_i$ and $1-\hat{p}_i$ is too small, we collect $\mathbf{f}^{(0)}_{i}$ into the hard instance set $\mathbf{T}$ as follows: 

\begin{equation}
\label{eq30}
\mathbf{T}=\left\{\mathbf{f}^{(0)}_i \mid \left|\hat{p}_i-\left(1-\hat{p}_i\right)\right|<\alpha\right\}_{i=1}^N,
\end{equation}
where $ i \in \{1,2,\ldots,N\}$ and $\alpha$ is the hard threshold.

When the number of hard instances in the WSI is greater than $\beta$ (threshold for the number of hard instances), we focus on mining the intrinsic difference in the hard instance set to distinguish task-specific and task-irrelevant features. We assign pseudo labels to these hard instances in $\mathbf{T}$ via unsupervised clustering, \emph{e.g.}, K-means algorithm. After clustering these hard instances into two clusters, we use the cluster label ${c}_i \in \{0,1\}$ to supervise the classification head $\mathcal{C}$ to learn from these hard instances. \textcolor{blue}{To ensure the cluster labels generated by unsupervised clustering are task-relevant, we employ a prediction-guided alignment strategy. For hard instances derived from negative bags, we assign a negative label ($c_i=0$) to both two clusters $\mathcal{U}_0$ and $\mathcal{U}_1$ as they are guaranteed to be negative. For hard instances derived from positive bags, we calculate the average prediction score $\bar{p}_k = \frac{1}{|\mathcal{U}_k|} \sum_{\mathbf{f}^{(0)}_i \in \mathcal{U}_k} \hat{p}_i$ for each cluster $k \in \{0,1\}$, where $\hat{p}_i$ is the prediction logit of feature $\mathbf{f}^{(0)}_i$. If $\bar{p}_0 < \bar{p}_1$, we assign the pseudo-label 0 to instances in $\mathcal{U}_0$ and 1 to those in $\mathcal{U}_1$, and vice versa. This strategy leverages the collective confidence of the classifier to correct individual noisy predictions.} This unsupervised mining can prevent the classifier from overfitting task-irrelevant features. 

The debiasing loss can be formulated as: 
\begin{equation}
\label{eq20}
\mathcal{L}_{debias}=\frac{1}{t}\sum_i^{t} \mathrm{CELoss}\left(\hat{{p}}_i,{c}_i\right),
\end{equation}
where $t$ is the number of hard instances in $\mathbf{T}$. The total loss is calculated as:
\begin{equation}
\label{eq21}
\mathcal{L}=\mathcal{L}_{cls}+\lambda \cdot \mathcal{L}_{debias},
\end{equation}
where $\lambda$ is a weight in the loss function to balance the debiasing loss and classifying loss.
 \begin{table}[t!]
\centering
\renewcommand\arraystretch{0.5}
\caption{\textcolor{blue}{Ablation study of two key components on TCGA-CRC-MSI cohort. w/o Mining: All instances are fed into two ViT layers without dictionary-guided instance grouping. \textcolor{blue}{The symbol ``*'' indicates only inter-group interaction mining is conducted without intra-group interaction mining.} w/o Debiasing: Remove the hard-instance-assisted classifier debiasing module.}
}\label{tab3}
\setlength{\tabcolsep}{1mm}
\renewcommand\arraystretch{0.3}
\begin{tabular}{cc|cc}
\toprule
\textbf{Pathology Mining} & \textbf{Debiasing} & \textbf{TCGA-CRC-MSI} & \textbf{TCGA-STAD-MSI} \\
\midrule
$\times$ & $\times$ & 88.63 & 64.05 \\
$\times$ & \checkmark & 91.49 & 79.73 \\
\textcolor{blue}{*} & \textcolor{blue}{$\times$} & \textcolor{blue}{88.70} & \textcolor{blue}{64.32} \\
\checkmark  & $\times$ & 93.46 & 80.00 \\
\checkmark  & \checkmark & \textbf{96.70} & \textbf{81.73} \\
\bottomrule
\end{tabular}
\vspace{-0.5em}
\end{table}

\section{Experimental Results}
\subsection{Datasets}
\noindent{\textbf{TCGA-CRC dataset. }}  The Cancer Genome Atlas \textcolor{blue}{\cite{weinstein2013cancer}} colorectal cancer dataset (TCGA-CRC) \textcolor{blue}{\cite{tcga2012comprehensive}} includes two subtypes: microsatellite stable (MSS) and microsatellite instability (MSI), collectively referred to as the TCGA-CRC-MSI cohort, as well as BRAF mutation and non-BRAF mutation, collectively referred to as the TCGA-CRC-BRAF cohort. The TCGA-CRC-MSI cohort consists of 420 MSS slides and 62 MSI slides, while the TCGA-CRC-BRAF cohort contains 429 non-BRAF slides and 53 BRAF slides. Each cohort is randomly divided into training and testing sets in a 3:1 ratio, with 10\% of the training set further allocated as the validation set.

\noindent{\textbf{CPTAC-COAD dataset. }} The Clinical Proteomic Tumor Analysis Consortium colon adenocarcinoma dataset (CPTAC-COAD) \textcolor{blue}{\cite{cptac_coad_dataset}, hosted on The Cancer Imaging Archive (TCIA) \cite{clark2013cancer},} includes two cohorts: CPTAC-COAD-MSI and CPTAC-COAD-BRAF. The CPTAC-COAD-MSI cohort contains 81 MSS slides and 24 MSI slides, while the CPTAC-COAD-BRAF cohort consists of 16 BRAF slides and 90 non-BRAF slides. This dataset is randomly split into training, validation, and testing sets in a 3:1:1 ratio.

\noindent{\textbf{TCGA-STAD dataset. }} The TCGA stomach adenocarcinoma dataset (TCGA-STAD) \textcolor{blue}{\cite{tcga2014comprehensive}} includes 224 MSS slides and 60 MSI slides, collectively referred to as the TCGA-STAD-MSI cohort. Preprocessed patches from all slides, as provided in \cite{kather2019deep}, are utilized in this study. We follow the data split strategy in \cite{kather2019deep}, dividing the dataset into training and testing sets and reserving 10\% of the training set as the validation set.

\textcolor{blue}{All data (including histological images) from the TCGA database are available at \url{https://portal.gdc.cancer.gov/}. All data from the CPTAC cohort are available at \url{https://proteomic.datacommons.cancer.gov/}. All clinical data for patients in the TCGA and CPTAC cohorts are available at \url{https://cbioportal.org/}.}
\textcolor{blue}{We strictly conduct patient-level splitting across all datasets. For patients containing multiple slides in the raw cohorts, we select a single slide (typically the diagnostic slide, DX) for each patient to construct the experimental cohorts, ensuring no patient overlap across the subsets.}

\subsection{Evaluation Metrics}
\textcolor{blue}{For all cohorts, we evaluate the performance of different methods with four metrics, \emph{i.e.}, Area Under Receiver Operator Curve  Score (AUROC) (\%), Accuracy (\%), Area Under the Precision–Recall Curve (AUPRC) (\%) and Balanced Accuracy (\%).} \textcolor{blue}{We report 95\% Confidence Intervals (CIs) for all metrics using non-parametric patient-level stratified bootstrapping. Statistical significance between methods is assessed using paired bootstrap tests, and results are considered statistically significant at $p<0.05$. All resampling procedures are stratified at the patient level.}
\begin{table}[t!]
\centering
\renewcommand\arraystretch{1.3}
\caption{\textcolor{blue}{Ablation study of grouping strategies on the TCGA-CRC-MSI and TCGA-STAD-MSI cohorts. Rand.: Randomly dividing instances. K-means: Clustering instances using K-means. Text: using GPT-4 TO GENERATE AS MANY DESCRIPTIONS AS POSSIBLE, \emph{i.e.}, 15 descriptions and grouping instances into 15 groups following \cite{zhang2024prompting}. k-NN (dist): k-NN grouping based on Euclidean distance. k-NN (cos): k-NN grouping based on cosine similarity. Dic.: Dictionary-guided grouping. The best result for each cohort is \textbf{highlighted}.}}
\label{tab4}
\setlength{\tabcolsep}{0.5mm}
\renewcommand\arraystretch{0.8}
\begin{tabular}{c|cccccc}
\toprule[1pt]
\textbf{Cohort} & \textbf{Rand.} & \textbf{Text} & \textbf{K-means} & \textbf{k-NN (dist)} & \textbf{k-NN (cos)} & \textbf{Dic.} \\
\hline
TCGA-CRC-MSI  & 89.46 & 91.75 & 92.83 & 92.00 & 90.86 & \textbf{96.70} \\
TCGA-STAD-MSI & 78.97 & 78.70 & 75.62 & 65.73 & 66.49 & \textbf{81.73} \\
\bottomrule[1pt]
\end{tabular}
\end{table}


\begin{figure}[t!]
\centering
\includegraphics[width=0.41\textwidth]{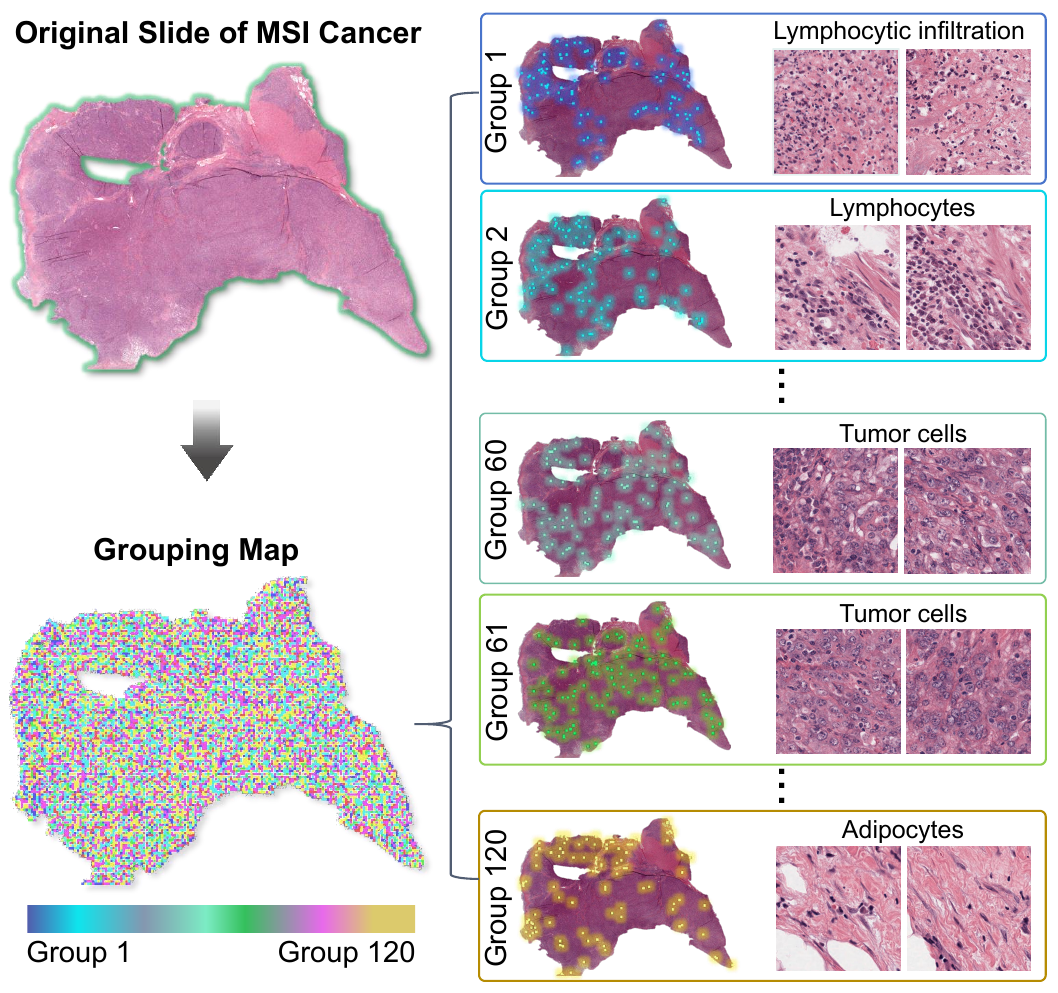}

\caption{Pathological group distribution and corresponding pathological patterns on a WSI of MSI cancer. 
} \label{groupmap}
\vspace{-0.5em}
\end{figure}

\subsection{Implementation Details}
\textcolor{blue}{To assess staining variability, we compute the inter-slide coefficient of variation (CV) based on the Optical Density (OD) of valid tissue regions on the training sets of each cohort. Specifically, tissue pixels are mapped to the OD space, strictly excluding background components using an OD threshold of 0.15. We then compute the slide-level mean OD and calculate the dataset-level CV across all WSIs. TCGA-CRC and TCGA-STAD exhibit substantial staining variation (CV=0.41 and 0.40, respectively), whereas CPTAC-COAD shows relatively low variation (CV=0.15).} For the TCGA-STAD-MSI cohort, we use preprocessed patches provided by \cite{kather2019deep}\textcolor{blue}{, where Macenko color normalization has already been applied.}. For other cohorts, we apply the image preprocessing algorithm from CLAM \cite{lu2021data} to detect tissue regions in WSIs and crop these regions into non-overlapping patches of size 512×512 pixels at 20× magnification. \textcolor{blue}{Given the high staining variability in TCGA-CRC, we apply Macenko color normalization \cite{macenko2009method} to this dataset. The target stain matrix is estimated from a single reference patch selected exclusively from the training set. Background and artifacts are excluded using brightness standardization and an OD threshold (\emph{i.e.}, $0.15$). The derived matrix is then applied to all training, validation, and test patches without using validation or test information. CPTAC-COAD is not normalized due to its relatively stable staining distribution.}

Following IBMIL \cite{lin2023interventional}, \textcolor{blue}{we utilize CtransPath (pretrained using self-supervised learning (SSL) on TCGA\footnote{https://portal.gdc.cancer.gov/} and PAIP\footnote{http://www.wisepaip.org/paip/} datasets) \cite{wang2022transformer} as the feature extractor for all experiments and additionally include UNI v2 \cite{chen2024towards} (pretrained on a massive dataset (Mass-100k [50]), distinct from TCGA) for the main comparative analysis}. \textcolor{blue}{Following standard protocols in WSI classification \cite{shao2021transmil, zhang2022dtfd}, we freeze these feature exractors. This choice is primarily driven by training efficiency and fair comparison with other methods.} Then we randomly sample 6000 instances in each WSI for training and testing. During training, the model is evaluated on the validation set after each epoch, and the parameters with the best performance are saved. \textcolor{blue}{To capture slide-specific hard instances, the K-means clustering is performed dynamically for each slide during the forward pass (\emph{i.e.}, per batch iteration) and the instance features fed into the clustering algorithm are detached from the computational graph.} We adopt the Adam optimizer using a learning rate of $5\times10^{{-5}}$ with learning rate annealing. Training is conducted for a maximum of 100 epochs, and performance is subsequently evaluated on the test set. The same experimental settings are applied across all competitors to ensure a fair comparison. 
All experiments are conducted on a single NVIDIA GeForce RTX 3090 GPU.


\subsection{Comparison with State-of-the-Art Methods}
We compare our framework on five cohorts with current state-of-the-art competitors: (1) ABMIL \cite{Ilse_Tomczak_Welling_2018}: a MIL method aggregating instance features via attention mechanism, (2) DSMIL \cite{li2021dual}: a MIL method aggregating instance features via non-local attention pooling, (3) CLAM \cite{lu2021data}: gated attention-base MIL method, (4) TransMIL \cite{shao2021transmil}: a MIL method aggregating instance features through MSA modules, (5) DTFD \cite{zhang2022dtfd}:  a MIL method introducing pseudo-bags and feature distillation, (6) HIPT \cite{chen2022scaling}: a hierarchical ViT framework leveraging Pyramid structure of WSIs, (7) MHIM-MIL \cite{tang2023multiple}: a MIL framework with masked hard instance mining, (8) WiKG \cite{li2024dynamic}: a graph-based method through knowledge-aware attention mechanism and (9) PromptBio \cite{zhang2024prompting}: our previous hierarchical MIL framework leveraging text to guide instance grouping. 
As shown in Table~\ref{tab1}, our method achieves the best performance on both evaluation metrics. Although WiKG \cite{li2024dynamic} and PromptBio \cite{zhang2024prompting} are comparable to our method in terms of Accuracy on the CPTAC-COAD-MSI cohort, they fall short in AUROC. This improvement is attributed to the learnable dictionary and hard-instance intervention, which enable the extraction of fine-grained pathological components and task-specific information. More detailed analysis is shown below. 
\begin{figure}[t!]
\centering
\includegraphics[width=0.49\textwidth]{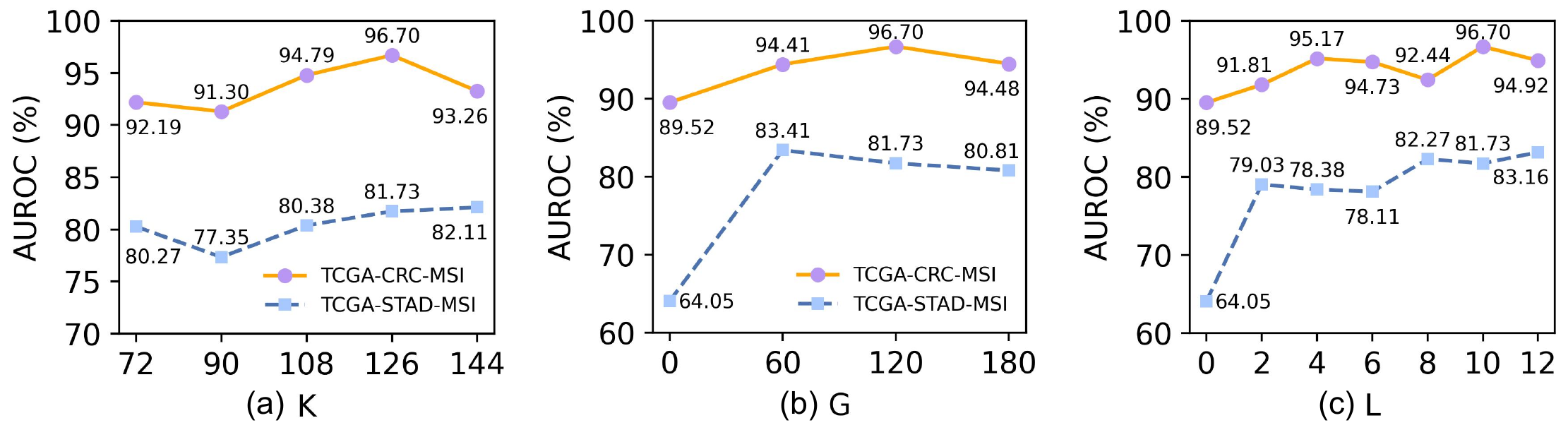}
\vspace{-1.6em}
\caption{\textcolor{blue}{Performance changes by varying the (a) dictionary size ($K$), (b) number of groups ($G$) and (c) number of dictionary updates ($L$) on the TCGA-CRC-MSI and TCGA-STAD-MSI cohort.}
} \label{fig9}
\vspace{-1.6em}
\end{figure}

\textcolor{blue}{Attention-based methods, \emph{e.g.}, ABMIL \cite{Ilse_Tomczak_Welling_2018}, CLAM \cite{lu2021data} and TransMIL \cite{shao2021transmil}, are limited in performance due to their focus on modeling relationships between instances.} A state-of-the-art graph-based method, WiKG \cite{li2024dynamic}, leverages learnable embeddings to capture instance similarities and constructs a graph based on those similarities. It still relies on the relationships between instances. Focusing on relationships between instances leads to inferior performance of these models compared to ours. While our dictionary-based method focuses on grouping instances based on their similarity to learnable tokens in the dictionary, enabling the model to extract more fine-grained pathological information and capture complex pathological structures through the dictionary. This enables our model to achieve the best performance. Our previous work, PromptBio \cite{zhang2024prompting}, which prompts a large language model to generate multiple descriptions for guiding pathological grouping, has limited performance due to the constraints of text descriptions.

To prevent the model from overfitting simple instances, MHIM-MIL \cite{tang2023multiple} mines hard instances by masking some high-attention instances, which results in the loss of important information, especially when the attention mechanism fails to capture task-relevant features. Thus, it leads to a drastic performance drop compared to our method. In contrast, our method drives the identification of hard instances directly based on the prediction confidence of the classifier, ensuring that these instances are challenging for the task. Moreover, all instance information is fully retained, and only hard instances are given more focus. This strategy avoids the exclusion of critical information while strengthening training in hard instances, allowing the model to capture task-relevant features.

Furthermore, as illustrated in Fig.~\ref{fig5}, our method identifies distinct pathological patterns in MSI and MSS cancers. MSI cancer is characterized by the presence of aggregated lymphocytes and lymphocytic infiltration, which indicate a stronger immune response, as well as aggregated tumor cells, cellular debris, and necrosis, reflecting rapid tumor mutation rates and growth. In contrast, MSS cancer exhibits sparse lymphocytes and plasma cells, associated with a weaker immune response.

\begin{figure}[t!]
\centering
\includegraphics[width=0.5\textwidth]{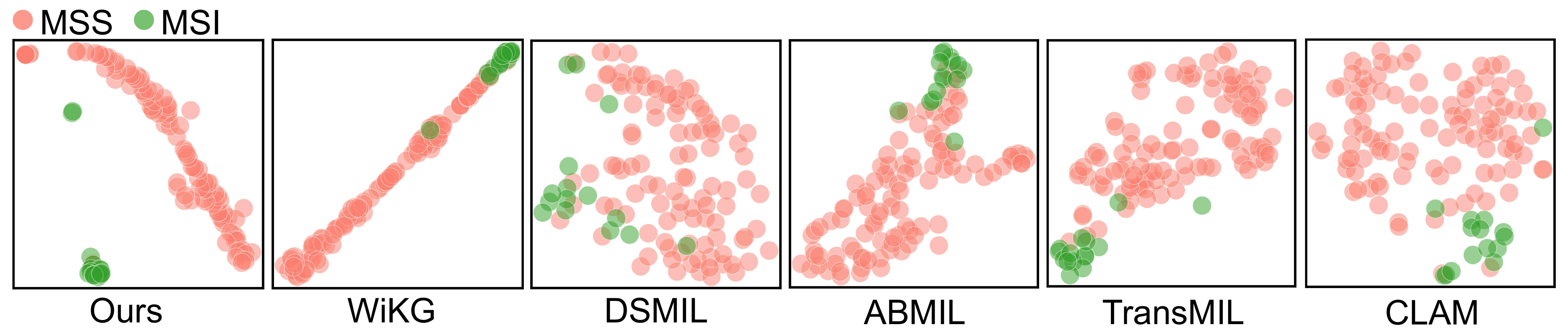}
\vspace{-1.6em}
\caption{t-SNE visualization of bag representations in different methods on the testing set of TCGA-CRC-MSI cohort. Our dictionary-based method constructs more discriminative representations of WSIs.
} \label{fig4}
\end{figure}
\begin{figure}[t!]
\centering
\includegraphics[width=0.5\textwidth]{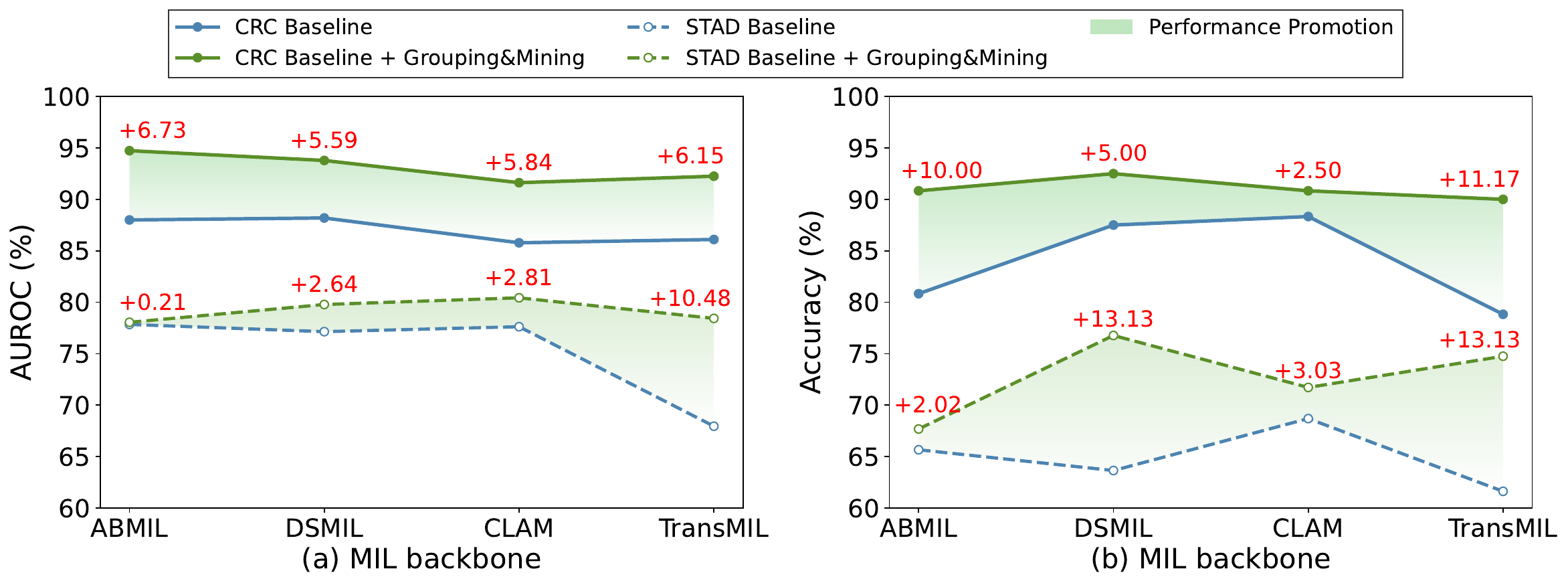}
\vspace{-1.6em}
\caption{\textcolor{blue}{(a) AUROC and (b) accuracy improvement on the TCGA-CRC-MSI cohort after employing dictionary-based hierarchical pathology mining to four MIL baselines, \emph{i.e.}, ABMIL \cite{Ilse_Tomczak_Welling_2018}, DSMIL \cite{li2021dual}, CLAM \cite{lu2021data} and TransMIL \cite{shao2021transmil}. Performance improvement is marked {\color{red}{red}}.} } \label{fig7}
\vspace{-1.6em}
\end{figure}
\subsection{Ablation Study}
We conduct a series of ablation studies to evaluate the effectiveness of key components in D$^2$Bio. These include: (1) the dictionary-based hierarchical pathology mining strategy, which encompasses grouping strategy, dictionary size ($K$), number of groups ($G$), number of dictionary updates ($L$) and its performance when applied to other MIL methods; (2) the hard-instance-assisted classifier debiasing strategy, considering factors such as the hard threshold ($\alpha$ in Eq.~\ref{eq30}), threshold for the number of hard instances ($\beta$), weight in loss function ($\lambda$ in Eq.~\ref{eq21}) and its performance when integrated with other MIL methods; and (3) the impact of different feature extractors. These experiments demonstrate the significance of each design choice in enhancing the overall performance of D$^2$Bio.
\vspace{5pt}

\noindent \textbf{(1) Ablation on key components.}

Table~\ref{tab3} summarizes the results of ablation study on two proposed modules. A significant performance drop without dictionary-based hierarchical pathology mining module can be observed (the $2$nd row in results), \emph{e.g.}, from 96.70\% to 91.49\% on the TCGA-CRC-MSI cohort, since it lacks of extracting fine-grained pathological components, struggling to represent the complex tumor micro-environment in WSIs. \textcolor{blue}{To assess whether modeling inter-pathological component interactions alone is sufficient, we conduct a variant (row 3 in Table~\ref{tab3}) without intra-group MSA. Performance drops markedly compared to both intra- and inter-pathological component interactions modeling (row 4), from 93.46\% to 88.70\% on TCGA-CRC and from 80.00\% to 64.32\% on TCGA-STAD, indicating that inter-group interactions cannot capture intra-group relationships. Both local and global interactions are required for accurate biomarker prediction.} Similarly, the absence of the hard-instance-assisted classifier debiasing module leads to a performance reduction (the $3$rd row in results), as the model tends to overfit simple features while failing to learn hard but task-relevant features.

\vspace{5pt}

\noindent\textbf{(2) Ablation on dictionary-based hierarchical pathology mining.}

\noindent\textbf{(i) Grouping strategy. }We study the effect of different grouping strategies, including random instance grouping, text-guided instance grouping \cite{zhang2024prompting}, K-means-guided instance grouping \cite{qu2022dgmil} and our dictionary-guided instance grouping, as depicted in Table~\ref{tab4}. Dictionary-guided instance grouping consistently outperforms other alternatives. This superiority can be attributed to its ability to extract meaningful pathological information from WSIs and achieve sufficiently fine-grained grouping, which cannot be obtained by other strategies. \textcolor{blue}{To validate the advantage of our dictionary-based grouping in modeling long-range dependencies, we compare it with two conventional dynamic edge construction algorithms based on k-nearest neighbor (k-NN): k-NN (dist) (k-NN grouping based on Euclidean distance) and k-NN (cos) (k-NN grouping based on cosine similarity). As shown in Table~\ref{tab4}, our dictionary-based approach significantly outperforms k-NN baselines (\emph{e.g.}, +4.7\% AUROC on TCGA-CRC). While k-NN methods tend to cluster spatially adjacent or locally similar patches, our learnable dictionary aggregates semantically consistent instances from across the entire WSI, effectively capturing global pathological contexts without spatial distance limitations.}

\begin{figure}[t!]
\centering
\includegraphics[width=0.48\textwidth]{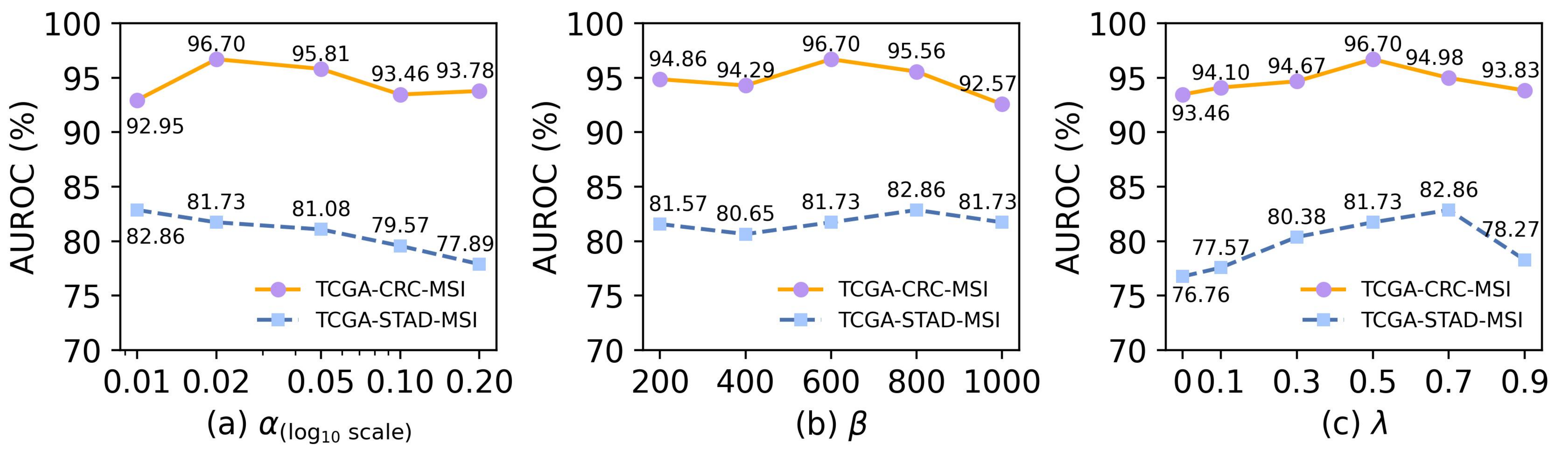}
\vspace{-0.6em}
\caption{\textcolor{blue}{Performance changes by varying the (a) hard threshold ($\alpha$), (b) threshold for the number of hard instances ($\beta$) and (c) weight in loss function ($\lambda$) on the TCGA-CRC-MSI and the TCGA-STAD-MSI cohort.} 
} \label{classifier_abltion}
\end{figure}

\begin{figure}[t!]
\centering
\includegraphics[width=0.48\textwidth]{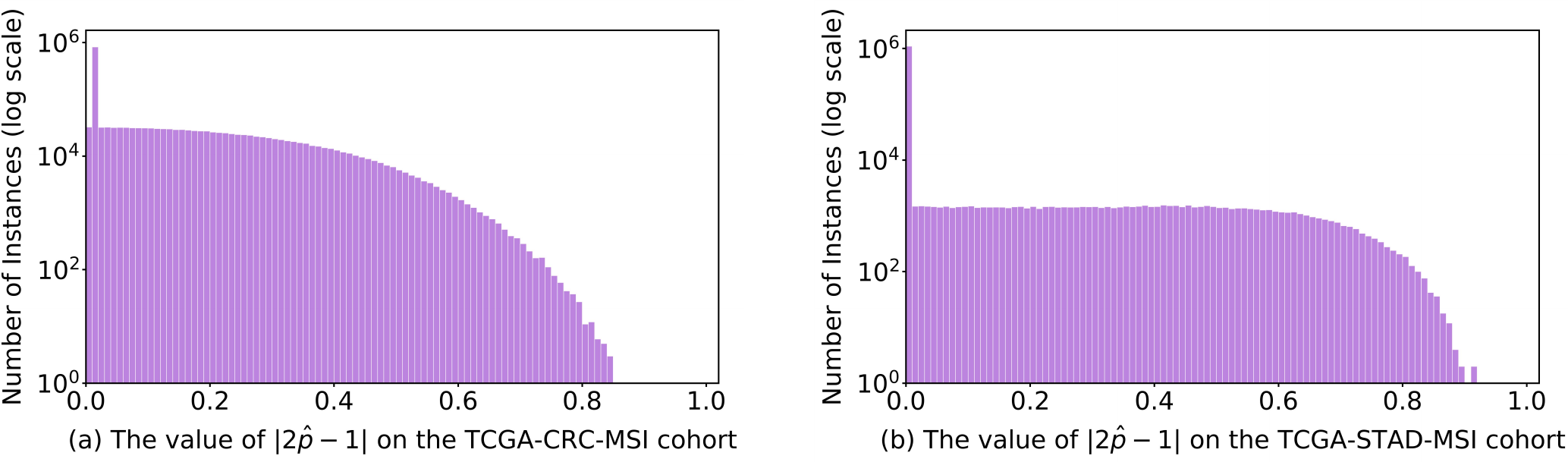}
\vspace{-0.6em}
\caption{\textcolor{blue}{Distributions of $|2\hat{p}_{i}-1|$ on the (a) TCGA-CRC-MSI and (b) TCGA-STAD-MSI cohort.} 
} \label{logit}
\end{figure}

\begin{figure}[t!]
\centering
\includegraphics[width=0.45\textwidth]{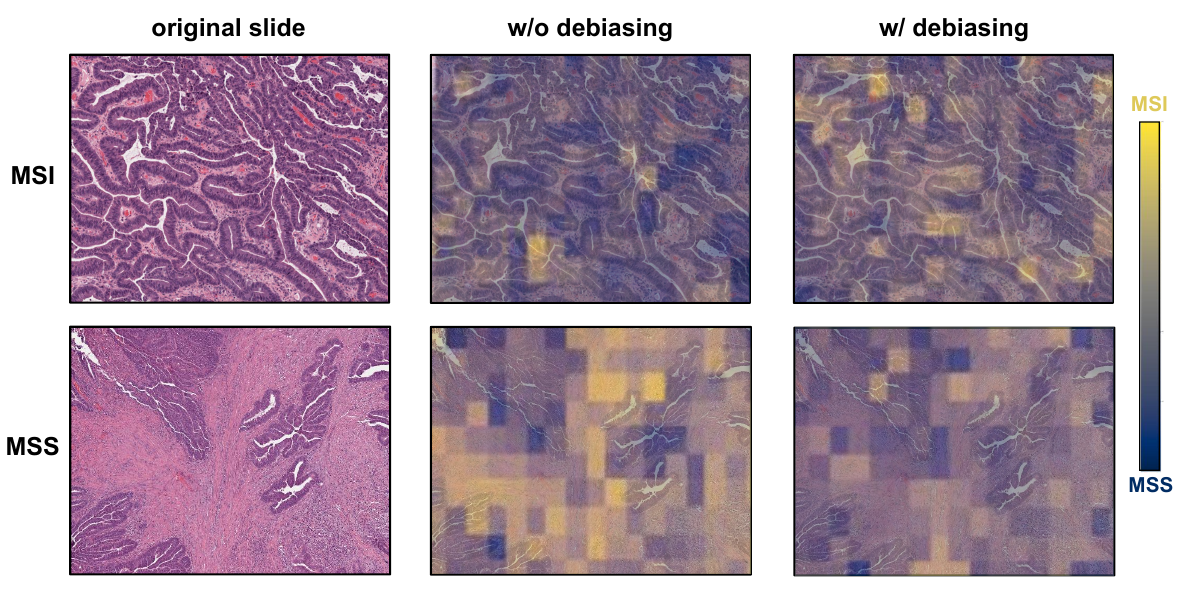}
\vspace{-0.6em}
\caption{Heatmap visualizations of D$^2$Bio w/o or w/ the hard-instance-assisted classifier debiasing module.
} \label{fig10}
\vspace{-1.6em}
\end{figure}
Fig.~\ref{groupmap} shows the distribution of extracted pathological groups on a WSI of MSI cancer. Even with long spatial distances, similar pathological patches are grouped together, \textcolor{blue}{verified by a clinical expert,} demonstrating our dictionary-guided instance grouping strategy effectively captures diverse pathological patterns. For example, patterns with lymphocytic infiltration and patterns with lymphocytes are grouped into Group 1 and Group 2, respectively, while patterns with adipocytes are grouped into Group 120. This shows dictionary-guided instance grouping is not limited to patch locations. Instead, it enables the model to capture diverse pathological features from WSIs. \textcolor{blue}{To ensure the stability of the learned representation, we evaluate the repeatability of the dictionary across independent runs. We observe that patches with similar morphological patches are consistently assigned to the same group, achieving a mean cosine similarity of 0.82 between matched atoms from different runs. This consistency allows the model to provide a stable and interpretable organization of the WSI components.}


\noindent\textcolor{blue}{\textbf{(ii) Dictionary size. } $K=126$ is set as the default value. We examine the impact of varying $K$ on both TCGA-CRC-MSI and TCGA-STAD-MSI cohorts, as shown in Fig.~\ref{classifier_abltion} (a). A consistent trend is observed across both cancer types: performance generally improves as $K$ increases in the range of $[108, 126]$. A smaller dictionary ($K<90$) leads to a noticeable performance drop, indicating that a sufficient number of atoms is essential to capture the diverse fine-grained pathological information.}

\noindent\textcolor{blue}{\textbf{(iii) Number of groups. } We set the number of groups $G=120$ by default. We further explore the impact of varying $G$ in Fig.~\ref{classifier_abltion} (b). While the optimal $G$ varies slightly due to biological heterogeneity (e.g., peaking at $G=60$ for STAD and $G=120$ for CRC), our default setting ($G=120$) maintains robust performance across cohorts, significantly outperforming the baseline without grouping ($G=0$). This suggests that dividing instances into approximately 120 groups provides a reliable grouping size for capturing semantic features across different cancer types.}

\noindent\textcolor{blue}{\textbf{(iv) Number of dictionary updates. } We set $L=10$ in the experiment. Fig.~\ref{classifier_abltion} (c) illustrates how performance varies with the number of dictionary updates. The model stabilizes as $L$ changes from 8 to 12 on both datasets. Setting $L=0$ (removal of dictionary-guided grouping) results in the lowest performance. The consistent stability after $L=8$ confirms that $L=10$ is a safe and robust choice for sufficient dictionary refinement.}




\noindent\textbf{(v) Application to other MIL methods. }Fig.~\ref{fig4} shows the feature mixtureness of bag representations on the testing set of TCGA-CRC-MSI cohort. Compared with other methods, our dictionary-based hierarchical pathology mining strategy yields more discriminative representations of WSIs. To study the effect of dictionary-based hierarchical pathology mining strategy, we employ it to four MIL backbones, as illustrated in Fig.~\ref{fig7}. By retaining dictionary-guided instance grouping and intra-group MSA, and replacing the inter-group ViT with these MIL backbones, significant performance improvement can be seen on all the four backbones. Notably, ABMIL \cite{Ilse_Tomczak_Welling_2018} and TransMIL \cite{shao2021transmil} benefit the most. Since the dictionary-based hierarchical pathology mining strategy enables these methods to focus on extracting interactions among pathological components rather than simple instance-level interactions.

\noindent\textcolor{blue}{\textbf{(vi) Impact of Instance Regrouping.} A potential concern with flattening and redividing instances is the disruption of initial cluster boundaries. We clarify that this design is a necessary trade-off to avoid CUDA OOM errors on high-end GPUs caused by the quadratic complexity of natural grouping on large WSIs. Furthermore, the iterative dictionary update acts as a dynamic routing mechanism, allowing instances to correct their group assignments over layers $L$. As visualized in Fig. 5, the final pathological groups exhibit strong biological consistency (e.g., separating tumor nests from stroma), confirming that the semantic structure is preserved and refined despite the intermediate redivision.}

\vspace{5pt}

\noindent\textbf{(3) Ablation on hard-instance-assisted classifier debiasing.}

\noindent\textbf{(i) Hard threshold.} The default hard threshold parameter is set to $\alpha=0.02$. \textcolor{blue}{To verify its generalization, we conduct sensitivity analyses, as shown in Fig.~\ref{classifier_abltion} (a). We observe that while the optimal $\alpha$ varies slightly (0.02 for CRC vs. 0.01 for STAD), the overall trend remains consistent: performance peaks at lower $\alpha$ values and consistently declines as $\alpha$ increases (\emph{e.g.}, beyond 0.05). To justify the choice of the uncertainty threshold, we analyze the distribution of $|2\hat{p}_{i}-1|$ for all instances in the training data. As shown in Fig.~\ref{logit}, most instances fall in the low-value range, representing high uncertainty. Our default setting $\alpha=0.02$ specifically targets these hard instances.} 

\noindent\textbf{(ii) Threshold for the number of hard instances. }The default threshold for the number of hard instances is $\beta=600$. \textcolor{blue}{As shown in Fig.~\ref{classifier_abltion} (b), the performance of D$^2$Bio is remarkably stable across this range for both cohorts. The optimal range for $\beta$ is found to be between 600 and 800. We selected $\beta=600$ as the default, as it consistently yields high AUROC while maintaining a sufficient number of hard patches for stable clustering across different WSIs.}

\noindent\textbf{(iii) Weight in loss function. }$\lambda=0.5$ is set as the default value. \textcolor{blue}{Fig.~\ref{classifier_abltion}} (c) demonstrates the impact of $\lambda$. Setting $\lambda=0$ means removal of the debiasing module. \textcolor{blue}{We observe that performance is not sensitive when $\lambda$ changes from 0.3 to 0.7 for both TCGA-CRC-MSI and TCGA-STAD-MSI cohort.} As $\lambda$ continues to increase, performance begins to decline. This is due to the loss of useful WSI labeling information, which is critical for accurate genetic biomarker prediction.

\begin{figure}[t!]
\centering
\includegraphics[width=0.49\textwidth]{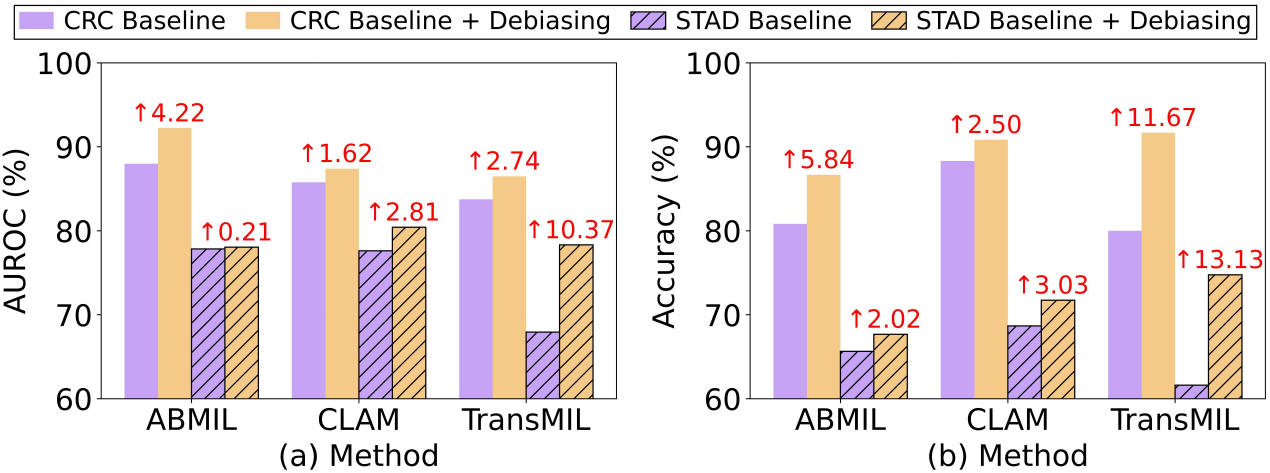}
\vspace{-1.6em}
\caption{\textcolor{blue}{(a) AUROC and (b) accuracy improvement on the TCGA-CRC-MSI cohort by employing hard-instance-assisted debiasing module to three MIL baselines: ABMIL \cite{Ilse_Tomczak_Welling_2018}, CLAM \cite{lu2021data} and TransMIL \cite{shao2021transmil}. Performance improvement is marked {\color{red}{red}}.} } \label{fig11}
\vspace{-0.6em}
\end{figure}


\begin{table*}[t]
\centering
\caption{\textcolor{blue}{Performance of DSMIL\cite{li2021dual}, WiKG\cite{li2024dynamic} and our D$^2$Bio using different pre-trained feature extractors on TCGA-CRC-MSI cohort and TCGA-SATD-MSI chohort. CtransPath \cite{wang2022transformer}: Trained on over 32,000 WSIs of TCGA and PAIP datasets. Virchow2 \cite{zimmermann2024virchow2}: Trained on 3.1 million WSIs from globally diverse institutions. UNI \cite{chen2024towards}: Trained on approximately 100,000 WSIs of the Mass-100k dataset. UNI v2 \cite{chen2024towards}: The extended version of UNI \cite{chen2024towards}, trained on approximately 350,000 WSIs. Best results are in \textbf{bold} and second-best results are \underline{underlined}.}} \label{extractor}
\setlength{\aboverulesep}{0.1pt}
\setlength{\belowrulesep}{0.1pt}
\setlength{\extrarowheight}{-1pt}
\renewcommand{\arraystretch}{0.8}
\setlength{\tabcolsep}{4pt}

\begin{tabular}{l|cccccc|cccccc}
\toprule
\multirow{3}{*}{Extractor} 
& \multicolumn{6}{c|}{TCGA-CRC-MSI} 
& \multicolumn{6}{c}{TCGA-STAD-MSI} \\

\cline{2-13}

& \multicolumn{2}{c}{DSMIL\cite{li2021dual}} 
& \multicolumn{2}{c}{WiKG\cite{li2024dynamic}} 
& \multicolumn{2}{c|}{D$^2$Bio (ours)} 
& \multicolumn{2}{c}{DSMIL\cite{li2021dual}} 
& \multicolumn{2}{c}{WiKG\cite{li2024dynamic}} 
& \multicolumn{2}{c}{D$^2$Bio (ours)} \\

\cline{2-13}

& AUROC & Acc. & AUROC & Acc. & AUROC & Acc. 
& AUROC & Acc. & AUROC & Acc. & AUROC & Acc. \\

\midrule
CtransPath \cite{wang2022transformer}    
& 88.19 & 87.50 & 90.22 & 90.00 & 96.70 & 92.50
& 77.14 & 63.64 & 79.84 & 70.71 & 81.73 & 77.78 \\

Virchow2 \cite{zimmermann2024virchow2}         
& 91.49 & 91.67 & 93.10 & 90.83 & \underline{97.52} & 93.33
& 84.22 & 78.79 & 82.97 & 77.78 & \textbf{86.65} & \textbf{83.84} \\

UNI \cite{chen2024towards}              
& 90.54 & 89.17 & 90.92 & 90.83 & 94.35 & \underline{94.17}
& 65.46 & 72.73 & 82.49 & 73.74 & \underline{86.32} & 74.75 \\

UNI v2 \cite{chen2024towards}                
& 89.21 & 90.83 & 95.49 & 94.10 & \textbf{98.48} & \textbf{96.67}
& 84.05 & 74.75 & 84.22 & 72.73 & 86.16 & \underline{82.83} \\

\bottomrule
\end{tabular}
\end{table*}

\begin{figure}[htbp]
\centering
\includegraphics[width=0.46\textwidth]{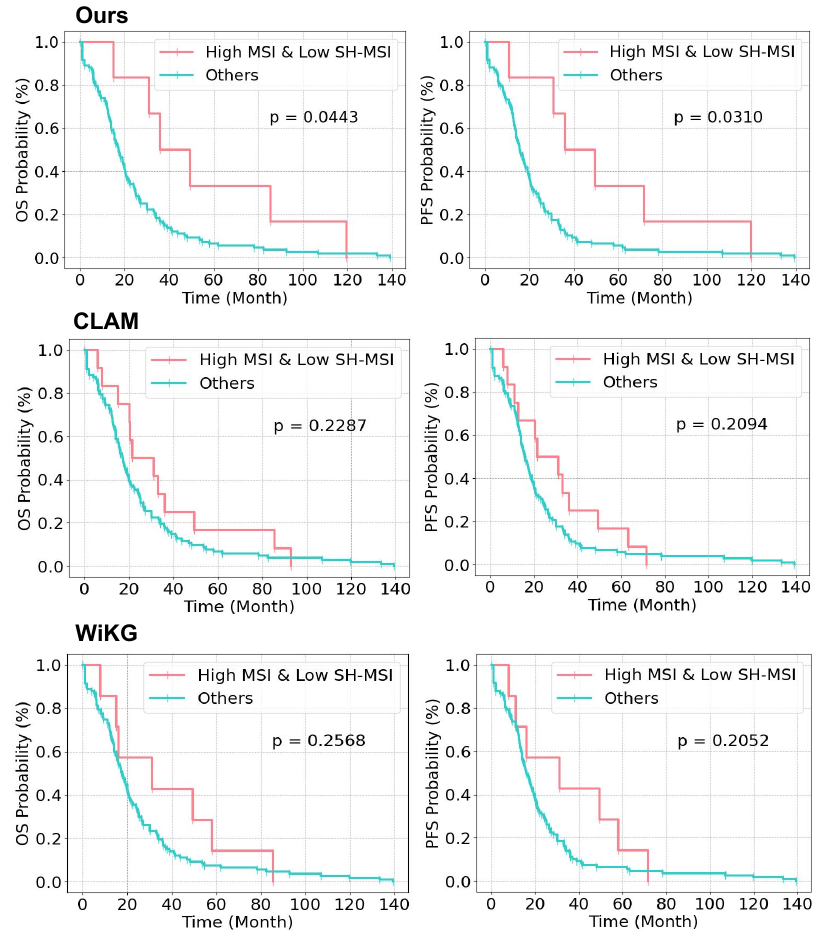}
\vspace{-0.6em}

\caption{KM plot of OS and PFS according to three methods, \emph{i.e.}, our D$^2$Bio, CLAM \cite{lu2021data} and WiKG \cite{li2024dynamic}, predicted patient-level ``High MSI \& Low SH-MSI" vs other subtypes. ``SH-MSI" refers to spatial heterogeneity of MSI.} \label{fig12}
\vspace{-1.6em}
\end{figure}






\noindent\textbf{(iv) Application to other MIL methods. }We seperately visualize the prediction heatmap of D$^2$Bio without or with the debiasing module in \textcolor{blue}{Fig.~\ref{fig10}}. Without the debiasing module, the model wrongly captures the features of normal stroma regions in the slide of MSS cancer, recognizing it as MSI cancer. With the debiasing module, the model shifts its focus more towards tumor regions and less on the stroma, thus recognizing more discriminative patterns that are relevant to the label. Considering the proposed hard-instance-assisted classifier debiasing is a plug-and-play module, we further carry out experiments to explore the influence of this strategy on three MIL baselines: ABMIL \cite{Ilse_Tomczak_Welling_2018}, CLAM \cite{lu2021data}, TransMIL \cite{shao2021transmil}, as illustrated in Fig.~\ref{fig11}. The results show that incorporating the hard-instance-assisted classifier debiasing module leads to performance improvements across all three MIL baselines. This is because the model is able to capture hard, task-related features that were previously overlooked.
\vspace{5pt}

\noindent\textbf{(4) Ablation on feature extractors. }


\textcolor{blue}{    To evaluate the robustness of D$^2$Bio, we conduct experiments using four feature extractors (CtransPath \cite{wang2022transformer}, Virchow2 \cite{zimmermann2024virchow2}, UNI and UNI v2 \cite{chen2024towards}) across two cohorts. As shown in Table~\ref{extractor}, our model consistently achieves the best performance using different extractors. Notably, UNI v2 \cite{chen2024towards} and Virchow2 \cite{zimmermann2024virchow2} further enhance the performance on CRC and STAD cohorts, respectively. These results indicate that D$^2$Bio works effectively with diverse advanced feature extractors, and stronger feature extractors consistently yield better performance.}

\noindent\textcolor{blue}{\textbf{(5) Computational Efficiency.}}

\textcolor{blue}{We further analyze the computational cost of D$^2$Bio in Table \ref{time}. On the TCGA-CRC-MSI cohort, our method requires approximately 60 seconds per epoch (using a single RTX 3090 GPU), compared to $\sim$9s for ABMIL, $\sim$10s for CLAM-SB and $\sim$27s for TransMIL. The increased time is attributed to the iterative dictionary updates and the per-bag dynamic clustering. However, the total training time for 100 epochs is approximately 1.7 hours, which remains highly efficient for training scenarios given the significant performance improvements.}

\begin{table}[h]
\centering
\caption{Average training time per epoch on TCGA-CRC-MSI.}
\label{time}
\begin{tabular}{l|c}
\toprule
\textbf{Method} & \textbf{Time per Epoch} \\
\midrule
CLAM-SB \cite{lu2021data} & $\sim$9 s \\
TransMIL \cite{shao2021transmil} & $\sim$10 s \\
ABMIL \cite{Ilse_Tomczak_Welling_2018} & $\sim$27 s \\
\textbf{D$^2$Bio (Ours)} & $\sim$60 s \\
\bottomrule
\end{tabular}
\end{table}

\subsection{Survival Analysis by MSI Prediction} 
Survival analysis is a crucial topic in clinical prognosis research, which aims to predict the time elapsed from a known origin to an event of interest, such as death and relapse of disease. We futher investigate whether patient-level MSI prediction generated by our D$^2$Bio could be useful to identify patient subgroups with distinct survival outcome on the testing set of TCGA-CRC-MSI cohort.

\textcolor{blue}{To maintain a consistent preprocessing pipeline across cancer types, we use the PLIP foundation model \cite{huang2023visual} for zero-shot tumor identification with two fixed prompts: ``an H\&E image of normal'' and ``an H\&E image of tumor''. Tiles are classified by argmax ($p_{tumor} > p_{normal}$) without cohort-specific tuning. An experienced pathologist independently review 500 predicted tumor and 500 predicted normal tiles from the CRC cohorts, yielding a precision of 92.2\% and confirming reliability for downstream analysis.} We first use PLIP \cite{huang2023visual} to select patches of tumor and follow \cite{xu2022spatial} to calculate the spatial heterogeneity of MSI (SH-MSI) of a WSI:
\begin{align}
&P_k = \hat{p}_k / N_{tumor},\\
&S=-\sum_k^N P_k \log _2\left(P_k\right),
\end{align}
where  $\hat{p}_k$ is the predicted MSI probability of $k$-th patch, $N_{tumor}$ is the number of tumor patches in the WSI and $N$ is the number of patches in  the WSI. 

\textcolor{blue}{Consistent with genomic biomarker studies \cite{davoli2017tumor, malta2018machine}, we utilize the average SH-MSI values of the testing set as the cutoff to determine high or low SH-MSI. Inspired by prior findings \cite{xu2022spatial} which demonstrates that patients with `High TMB \& Low SH-TMB' exhibit the best survival outcomes compared to others, and given the strong biological correlation between Tumor Mutation Burden (TMB) and MSI \cite{chalmers2017analysis}, we hypothesize that `High MSI \& Low SH-MSI' would similarly represent the best prognostic subgroup in colorectal cancer. Therefore, we divide testing set into two subtypes, \emph{i.e.}, ``High MSI \& Low SH-MSI" and ``Others''} The experiment is \textcolor{blue}{conducted} on three models respectively, \emph{i.e.}, our D$^2$Bio, CLAM \cite{lu2021data} and WiKG \cite{li2024dynamic}. 

The Kaplan-Meier (KM) plot of overall survival (OS) and progression-free survival (PFS) shown in Fig.~\ref{fig12} indicates that the two subgroups based on our D$^2$Bio have statistically significant differences in OS and PFS ($p$-values of 0.0443 and 0.0310, respectively), while predictions from CLAM \cite{lu2021data} and WiKG \cite{li2024dynamic} fail to identify distinct survival outcomes among patient subgroups. The survival analysis highlights the clinical utility of our D$^2$Bio in integrating MSI and its spatial heterogeneity as a prognostic biomarker.

\FloatBarrier
\section{Conclusion}
In this work, we propose D$^2$Bio, a WSI-based genetic biomarker prediction framework, to tackle the problem of constructing a pathology-aware representation involving the complex interconnections among pathological components and preventing the model from overfitting simple but irrelative instances. 
The dictionary-based hierarchical pathology mining module is proposed to mine diverse and very fine-grained pathological contextual interaction among various components in WSIs, without the limit to the distances between patches. Furthermore, hard-instance-assisted classifier debiasing module is designed to learn a debiased classifier via focusing on hard but task-related features. Experimental results on five cohorts demonstrate that our model significantly outperforms other state-of-the-art methods. Our analysis highlights the clinical interpretability of D$^2$Bio in genetic biomarker diagnosis and its potential utility in survival analysis. Our proposed modules can be easily plugged into other MIL methods and can be further validated in future research studies.

\FloatBarrier
\bibliographystyle{IEEEtran}
\bibliography{main}
\end{document}